\relax
\documentclass{article}
\usepackage{fullpage}
\usepackage{times}  % DO NOT CHANGE THIS
\usepackage{helvet}  % DO NOT CHANGE THIS
\usepackage{courier}  % DO NOT CHANGE THIS
\usepackage[hyphens]{url}  % DO NOT CHANGE THIS
\usepackage{graphicx} % DO NOT CHANGE THIS
\usepackage[square]{natbib}  % DO NOT CHANGE THIS AND DO NOT ADD ANY OPTIONS TO IT
\usepackage{caption} % DO NOT CHANGE THIS AND DO NOT ADD ANY OPTIONS TO IT
\DeclareCaptionStyle{ruled}{labelfont=normalfont,labelsep=colon,strut=off} % DO NOT CHANGE THIS
\usepackage{algorithm}
\usepackage{algorithmic}
\usepackage[dvipsnames]{xcolor}

\usepackage{amsmath}
\usepackage{amssymb}
\usepackage{newfloat}
\usepackage{listings}
\usepackage{booktabs}
\usepackage{multirow}
\usepackage{subcaption}
\usepackage{fontawesome5}
\usepackage[export]{adjustbox}
\usepackage{hyperref}
\hypersetup{colorlinks,citecolor=red}
\definecolor{blue}{HTML}{15316E}
\definecolor{red}{HTML}{800000}

\newtheorem{theorem}{Theorem}

\newtheorem{proposition}[theorem]{Proposition}

\floatstyle{ruled}
\newfloat{listing}{tb}{lst}{}
\floatname{listing}{Listing}

\usepackage{float}

\title{\textbf{Leaping Through Time with Gradient-based Adaptation for Recommendation}}

\author{
  \textbf{Nuttapong Chairatanakul}\\
  Tokyo Tech \& RWBC-OIL, AIST \\
  Tokyo, Japan \\
  \texttt{nuttapong.c@net.c.titech.ac.jp}
  \and
  \textbf{Hoang NT} \\
  Tokyo Tech \\
    Tokyo, Japan \\
  \texttt{hoangnt@net.c.titech.ac.jp}
  \and
  \textbf{Xin Liu} \\
  AIRC, AIST \& RWBC-OIL, AIST \& DigiARC, AIST \\
    Tokyo, Japan \\
  \texttt{xin.liu@aist.go.jp}
  \and
  \textbf{Tsuyoshi Murata} \\
  Tokyo Tech \& RWBC-OIL, AIST \\
    Tokyo, Japan \\
  \texttt{murata@c.titech.ac.jp}
}

\date{}

\begin{document}

\maketitle

\begin{abstract}
Modern recommender systems are required to adapt to the change in user preferences and item popularity.
Such a problem is known as the temporal dynamics problem, and it is one of the main challenges in recommender system modeling.
Different from the popular recurrent modeling approach, we propose a new solution named LeapRec to the temporal dynamic problem by using trajectory-based meta-learning to model time dependencies.
LeapRec characterizes temporal dynamics by two complement components named global time leap (GTL) and ordered time leap (OTL).
By design, GTL learns long-term patterns by finding the shortest learning path across unordered temporal data.
Cooperatively, OTL learns short-term patterns by considering the sequential nature of the temporal data.
Our experimental results show that LeapRec consistently outperforms the state-of-the-art methods on several datasets and recommendation metrics.
Furthermore, we provide an empirical study of the interaction between GTL and OTL, showing the effects of long- and short-term modeling.
\end{abstract}

\newcommand{\red}[1]{{\color{red} #1}}
\newcommand{\needcite}{\red{[C]}}
\newcommand{\model}{{LeapRec}}
\renewcommand{\cite}{\citep}
\section{Introduction}

Recommender systems \citep{schaferRecommenderSystemsEcommerce1999,ricciIntroductionRecommenderSystems2011,Aggarwal2016} have become essential tools in most real-world applications of big data.
These systems ease the users' effort in selecting suitable \emph{items} for their needs (e.g., commercial products, books, services) by finding common patterns of their preferences.
In practice, the preferences of users and their opinions of items frequently change over time~\cite{korenCollaborativeFilteringTemporal2009,xiongTemporalCollaborativeFiltering2010}.
It poses a challenge in designing recommender systems: \emph{How to incorporate temporal dynamics into recommendation results?}

Following the footsteps of sequential neural modeling, recommender systems researchers have designed a wide range of recurrent neural networks (RNNs) to solve the temporal dynamic challenge~\citep{hidasiSessionbasedRecommendationsRecurrent2016,liNeuralAttentiveSessionbased2017,zhuWhatNextModeling2017,maMemoryAugmentedGraph2020}.
The main idea of these architectures is to capture the relationship between the current and the past interactions using recurrent units~\citep{wangSequentialRecommenderSystems2019a}.
However, applying RNNs to new types of learning models is often difficult~\citep{pascanuDifficultyTrainingRecurrent2013,yuReviewRecurrentNeural2019};
For example, the graph neural network (GNN) is a promising model that has recently become the state-of-the-art for recommendations on non-temporal datasets~\citep{wangNeuralGraphCollaborative2019,heLightGCNSimplifyingPowering2020}.
Applying the recurrent architecture to a GNN would require redesigning the whole neural network~\cite{xuInductiveRepresentationLearning2019,tgn_icml_grl2020}.
Therefore, an interesting research question arises: \emph{Is there a more versatile alternative solution than RNNs to model temporal dynamics?}

Drawing inspirations from the meta-learning literature, which studies fast adaptation to new tasks \citep{vilaltaPerspectiveViewSurvey2002}, 
we hypothesize that meta-learning methods could be useful to model temporal dynamics.
Research on meta-learning for differentiable models demonstrated the advantages of a gradient-based method named MAML~\citep{finnModelAgnosticMetaLearningFast2017} over RNN-based models~\citep{raviOptimizationModelFewShot2016,santoroMetaLearningMemoryAugmentedNeural2016} in the few-shot learning problem.
This technique leverages the gradient path of the optimization process to learn an initialization of parameters, which allows fast adaptation to unseen tasks.

\begin{figure}[ht]
    \centering
    \includegraphics[width=0.7\linewidth]{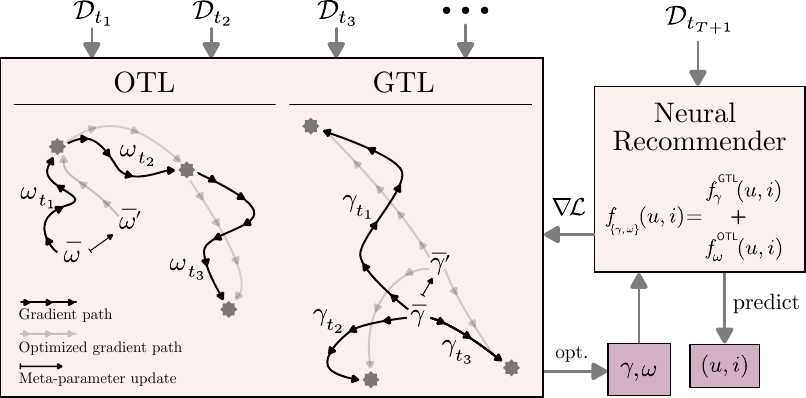}
    \caption{\emph{Left}: Our proposed LeapRec consists of GTL and OTL. \emph{Right}: A Neural Recommender. LeapRec enables any neural recommender to learn temporal dynamics by optimizing the gradient paths $\{\gamma_i\}_{i=t_1}^{t_T}$ and $\{\omega_j\}_{j=t_1}^{t_T}$.}
    \label{fig:overview}
\end{figure}

In this paper, we study the application of meta-learning to model the temporal dynamics in recommender systems.
The overall approach is to consider the recommendation problem on data from each time period as a separate task.
This approach implies that each time step is modeled as a sequence of gradient steps (a gradient path) in the optimization process.
However, a straightforward application of MAML poses two main issues.
First, since the computational cost of MAML mainly depends on the number of gradient steps and the number of gradient steps grows linearly with the number of time steps, 
it becomes infeasible to keep track of long-term time dependency.
Second, although we can reduce the computational cost using the first-order approximation of MAML (FOMAML)~\citep{finnModelAgnosticMetaLearningFast2017,nichol2018firstorder}, this only considers the start and end points of the gradient paths.
That is, FOMAML ignores the \emph{gradient trajectory}\footnote{Gradient trajectory is the set of directions in the gradient path.}, which leads to a loss in temporal information.
More recently, trajectory-aware meta-learning was proposed by~\citet{flennerhagTransferringKnowledgeLearning2019} to reduce the computational cost of MAML without sacrificing performance by minimizing the total length of gradient paths across tasks.
In the temporal recommendation context, we found that Leap is a promising approach because it considers the entire gradient trajectory while maintaining the computation at minimal cost.

We propose LeapRec as a novel adaptation of trajectory-based meta-learning to recommender systems.
Figure~\ref{fig:overview} illustrates an overview of our approach.
In LeapRec, we divide temporal dynamics into two categories: global time leap (GTL) and ordered time leap (OTL).
GTL finds common patterns that share across all time periods.
Examples of these patterns in e-commerce are classic books, basic goods, and all-time best seller items.
On the other hand, OTL captures the temporal dynamics from momentary social trends. 
Our experimental results show that our implementation of LeapRec recommendation model (Figure~\ref{fig:recommender}) outperforms the state-of-the-art (SOTA) methods on benchmark datasets by a large margin.
Furthermore, LeapRec, when used as a plug-in method for other recommender systems, also significantly improves their performances. The code and resources are available at \url{https://github.com/nutcrtnk/LeapRec}.
We summarize our contributions as follows.

\begin{itemize}
    \item To the extent of our knowledge, LeapRec is the first trajectory-based meta-learning approach to recommender systems.
    \item To fully utilize the trajectory-based approach to recommender systems, we propose two components of LeapRec named Global and Ordered Time Leap.
    \item Our empirical results show a clear improvement over existing approaches, and our empirical analyses explain the dynamic behavior of LeapRec. 
\end{itemize}

\section{Related Work}
\label{sec:related-work}

\subsection*{Sequential Recommendation}
Sequential history of user interactions provides valuable information about their preferences and dependencies between items.
Naturally, RNN-based models~\citep{hidasiSessionbasedRecommendationsRecurrent2016,liNeuralAttentiveSessionbased2017,wuRecurrentRecommenderNetworks2017,chairatanakulRecurrentTranslationBasedNetwork2019} have become the most popular choice for modern recommender systems, owing to their effectiveness in handling sequential data.
However, this approach is limited by the memory of RNNs: the amount of memory required to capture long-term dependencies grows linearly with the number of time steps.
Since it is infeasible to design such a learning model, \citet{chenSequentialRecommendationUser2018,liuSTAMPShortTermAttention2018} proposed the use of memory networks~\citep{westonMemoryNetworks2015} to improve long-term dependencies modeling in recommender systems.
% The history of interactions of each user provide crucial 
In parallel with these developments in RNNs, self-attention (SA) models~\citep{vaswaniAttentionAllYou2017} have gained popularity due to their success in natural language processing. 
Their popularity inspires researchers to investigate the applications of self-attention to improve recommendation quality~\citep{kangSelfAttentiveSequentialRecommendation2018,sunBERT4RecSequentialRecommendation2019}.
These models have the advantage that they can partially adapt to recent data based on the change in input sequences.
However, the dependencies between items can be outdated (e.g., combinations of fashion items), and the models cannot differentiate the outdated and new trends.

\subsection*{Temporal-aware Recommendation}
In the real world, the preference of users and their opinions of items change over time.
Temporal-aware recommendation models explicitly incorporate time into their recommendation results.
TimeSVD++~\cite{korenCollaborativeFilteringTemporal2009} is a pioneer work, which includes a time-bias factor.
\citet{xiongTemporalCollaborativeFiltering2010} proposed tensor factorization with time as a dimension.
Along with recent developments in the sequential recommendation, Time-LSTM~\citep{zhuWhatNextModeling2017}, MTAM~\citep{jiSequentialRecommenderTimeaware2020}, TiSASRec~\citep{liTimeIntervalAware2020}, and TGSRec~\citep{fanContinuousTimeSequentialRecommendation2021} incorporate time intervals between successive interactions into LSTM, Memory-Network, and SA;
SLi-Rec~\citep{yuAdaptiveUserModeling2019} and TASER~\citep{yeTimeMattersSequential2020} consider both time intervals and the time of the prediction. 
Each design has its own modifications that refine a few specific components to be time-aware.
Thus, it is difficult to transfer the design of temporal components to other architectures.

Another type of temporal-aware models is streaming or online recommendation~\citep{heFastMatrixFactorization2016,changStreamingRecommenderSystems2017,wangStreamingRankingBased2018}, where we can access only the most recent data but not the complete historical data. 
Training only on recent data inevitably shifts the model to be more biased toward such data than the past.
However, this poses a problem in retaining past information \citep{wangNeuralMemoryStreaming2018}.

\subsection*{Meta-learning in Recommender Systems}
With the concept of meta-learning that aims to make learning faster, \citet{vartakMetalearningPerspectiveColdstart2017,leeMeLUMetaLearnedUser2019,weiFastAdaptationColdStart2020,wangSequentialRecommendationColdstart2021} adopted meta-learning techniques to alleviate the \emph{cold-start} problem, where a model has not yet gathered sufficient information to draw any inferences for new users or new items.
\citet{liuIntentPreferenceDecoupling2020} applied meta-learning for session-based recommendation~\citep{hidasiSessionbasedRecommendationsRecurrent2016}. \citet{luoMetaSelectorMetaLearningRecommendation2020} considered an application of meta-learning to select a suitable model for each user. 
Most of these approaches are based on MAML, which has been shown to be effective for few-shot learning.

Meta-learning also has been applied to improve recommendation in online learning scenario. For example, \citet{zhangHowRetrainRecommender2020} proposed SML, a CNN-based \emph{meta-update} method that uses MAML for meta-training; \citet{xieLongShortTermTemporal2021} modified MAML for enhancing time adaptation of their recommender system.

\newcommand{\User}{\mathcal{U}}
\newcommand{\Item}{\mathcal{I}}
\newcommand{\Time}{\mathcal{T}}
\newcommand{\Seq}{\mathcal{S}}
\newcommand{\data}{\mathcal{D}}
\newcommand{\Graph}{{G}}
\newcommand{\Node}{\mathcal{V}}
\newcommand{\Edge}{\mathcal{E}}
\newcommand{\R}{\mathbb{R}}
\newcommand{\func}[1]{#1}
\newcommand{\num}[1]{{\vert#1\vert}}
\newcommand{\norm}[1]{{\Vert#1\Vert^2}}
\newcommand{\Norm}[1]{{\Big\Vert#1\Big\Vert^2}}
\newcommand{\loss}{\mathcal{L}}
\newcommand{\expect}{{\mathbb{E}}}
\newcommand{\emb}[1]{{\mathbf{#1}}}
\newcommand{\latestTime}{T}
\newcommand{\GT}{\gamma}
\newcommand{\OT}{\omega}
\newcommand{\Lparam}[3]{{#1}_{#2}^{#3}}
\newcommand{\GM}{\bar{\gamma}}
\newcommand{\OM}{\bar{\omega}}
\newcommand{\TT}{\theta}
\newcommand{\TM}{\bar{\theta}}

\section{Preliminaries} \label{sec:preliminaries}

\subsubsection*{Problem setup}
Let $\User = \{u_1, u_2, \ldots, u_U \}$ be a finite set of users, $\Item = \{i_1, i_2, \ldots, i_I \}$ be a finite set of items, and $\Time = (t_1, t_2, \ldots, t_T )$ be a finite sequence of timestamps. 
The cardinalities of these sets correspond to the number of users $U=\num{\User}$, number of items $I=\num{\Item}$, and number of timestamps $T=\num{\Time}$. 
Given sets of interactions between users and items in each timestamps:  $\data = \{ \data_{t_1}, \data_{t_2}, \ldots, \data_{t_T} \}$, where $t \in \Time \text{ and } \data_t \in 2^{\ \User \times \Item}$. 
The goal is to build a learning model to predict new interactions after $t_T$ with high accuracy.
In practice, such a model is evaluated using a test dataset $\{\data_{t_{T+1}}, \data_{t_{T+2}}, \ldots \}$.

\subsubsection*{Objective function for recommendation}
To model how likely a user $u \in \User$ would interact with an item $i \in \Item$, we define $f_\theta(u,i)$ to be a scoring function parameterized by $\theta$.
Commonly, Bayesian personalized ranking (BPR)~\citep{rendleBPRBayesianPersonalized2009} is a suitable choice for the optimization target.
The core idea is to maximize the interaction likelihood by using the contrastive learning principle.
The model $\func{f}_\theta$ is trained to differentiate between the rankings of observed (positive) and unobserved (negative) interactions.
Formally, we learn the recommendation model $\func{f}_\theta$ by minimizing the following BPR loss with respect to parameter $\theta$.
\begin{equation}
    \label{eq:loss}
    \loss(\theta; \data) = \mathop{\expect}_{(u,i) \sim \data, (u,j) \sim \tilde{\data}}[-\sigma\left(\func{f}_\theta (u,i) - \func{f}_\theta (u,j)\right)],
\end{equation}
where $\sim$ denotes i.i.d. uniform samplings, $\tilde{\data}$ is the negative of $\data$: $\tilde{\data} = (\User \times \Item)\setminus \data$, and $\sigma$ is the sigmoid function.
Throughout this paper, we use BPR as the loss function.

\newcommand{\update}[2]{\alpha \nabla \loss_{#2}({#1_{#2}}) }

\subsection*{Leap: Trajectory-based Meta-Learning}
\label{ssec:trajectory-meta-learning}

Gradient-based meta-learning aims to obtain an initialization of parameters, also named meta-parameters $\TM$, such that it allows fast adaptation across different learning tasks via the gradients~\citep{finnModelAgnosticMetaLearningFast2017}.
\citet{flennerhagMetaLearningWarpedGradient2019} recently proposed a trajectory-based meta-learning method that minimizes the length of the learning process.
The length of the learning process consists of (i) the distance between the loss landscape and (ii) the distance in the parameter space of an update, namely from $\TT_\tau^k$ to $\TT_\tau^{k+1}$ with $\TT^0_\tau \gets \TM$.
That results in the meta-gradient
\begin{equation} \label{eq:leap:meta_grad}
    \nabla \func{F}_{\TM} = \expect_{\tau \sim p(\tau)} \left[ \sum_{k=0}^{K-1} - ( \underbrace{ \Delta \loss^k_\tau \nabla \loss_\tau (\TT_\tau^k) }_{\text{(i)}} + \underbrace{ \Delta \TT^k_{\tau} }_{\text{(ii)}} ) \right],
\end{equation} 
where $\tau$ represents a task sampled from some distribution $p(\tau)$, $K$ is the number of update steps,
$\mathcal{L}_\tau$ is the loss with respect to $\tau$,
$\Delta \loss^k_\tau = \loss_\tau( \TT^{k+1}_\tau ) -  \loss_\tau ( \TT^k_\tau )$, and $\Delta \TT^k_{\tau} = \TT^{k+1}_{\tau} - \TT^{k}_{\tau}$.
Note that the Jacobian term is omitted from our equation because using the identity matrix should provide a good approximation~\citep{flennerhagMetaLearningWarpedGradient2019}.
Subsequently, we define the update of $\TT$ by adopting a simple gradient-based update, following MAML:
$\TT^{k+1}_{\tau} \gets \TT^k_\tau - \update{\TT^k}{\tau},$
where $\alpha$ denotes the learning rate.

\section{Proposed Method} \label{sec:method}

\subsection{Leaping Through Time}
\label{ssec:leap-time}

We adopt Leap to our recommendation models by considering data from each time step as a task.
This section proposes two distinct components of LeapRec named Global Time Leap (GTL) and Ordered Time Leap (OTL)\footnote{ We provide further analyses of GTL/OTL in the Appendix.}.
We denote $\GM$ and $\GT$ to be meta-parameters and model parameters of GTL while denoting $\OM$ and $\OT$ to be meta-parameters and model parameters of OTL.

\newcommand{\grad}[1]{\nabla \func{F}_{#1}}
\begin{algorithm}[h]
\caption{Meta-optimization for $\GM$ and $\OM$ using Leap}
\begin{algorithmic}[1]
    \REQUIRE $\alpha, \beta, \eta$: learning rate parameters
    \STATE initialize $\GM$ and $\OM$
    \WHILE{not done}
        \STATE $\grad{\GM} \gets 0$, $\grad{\OM} \gets 0, \, \OT' \gets \OM$
        \FORALL{$t \in \Time$}
            \STATE $\GT_t^0 \gets \GM$, $\Lparam{\OT}{t}{0} \gets \OT'$
            \FOR{$k \gets 0$ to $K-1$}
                \STATE $\GT^{k+1}_{t} \gets \GT^k_t - \alpha \nabla_{\GT_k^t} \loss_t ( {\{\GT^k_t, \Lparam{\OT}{t}{k} \}} )$
                \STATE $\Lparam{\OT}{t}{k+1} \gets \Lparam{\OT}{t}{k} - \alpha \nabla_{\Lparam{\OT}{t}{k}} \loss_t ( {\{\GT^k_t, \Lparam{\OT}{t}{k} \}})$
                % \STATE $\GM_t^k \gets \GT_t^k, \, \Lparam{\OM}{t}{k} \gets \Lparam{\OT}{t}{k}$

            \ENDFOR
            \STATE $\OT' \gets \Lparam{\OT}{t}{K}$
        \ENDFOR
        \STATE update $\grad{\GM}, \grad{\OM}$ using Eq. \eqref{eq:leap:meta_grad}.
        \STATE $\GM \gets \GM - \frac{\beta}{T} \grad{\GM}, \, \OM \gets \OM - \eta \grad{\OM}$
    \ENDWHILE
\end{algorithmic}
\label{alg:training}
\end{algorithm}

\subsubsection{Global Time Leap (GTL)}
GTL aims to find common patterns across all time steps by minimizing their gradient paths starting from $\GM$ (see Figure~\ref{fig:overview} for an illustration).
We define a \emph{task} as the recommendation objective under each timestamp with the following loss function: $\loss_t (\GT) = \loss(\GT; \data_t)$.
The gradient update can be defined as: $\GT^{k+1}_t \gets \GT^k_t - \update{\GT^k}{t}$ with $\GT_t^0 \gets \GM$.
Because the parameters of GTL will revert to $\GM$ that stores the information across all timestamps before each update, this mechanism imitates the \emph{static} or long-term preferences of users across all time steps.

\newcommand{\lossupdate}[1]{\Lparam{\OT}{#1}{0} - \alpha \sum_{k=0}^{K-1}  \nabla \loss_{#1}({\Lparam{\OT}{#1}{k}})}

\subsubsection{Ordered Time Leap (OTL)}
While it is straightforward to use meta-learning in this context by assuming each time step is independent, encoding the sequential nature of temporal data is a non-trivial problem.
Our idea is to capture the temporal information using the gradient trajectory that accumulates across the time sequence.
We denote $\Lparam{\OT}{t_i}{k}$ to be the value of $\OT$ at gradient step $k$ and time step $t_i$.
The meta-parameters $\OM$ serves as the starting point for $\Lparam{\OT}{t_1}{0}$.
We update the model parameters and accumulate the gradient along the trajectory as follows.
Within the same time step $t_i$, we update with learning rate $\alpha$: $\Lparam{\OT}{t_i}{K} \gets \lossupdate{t_i}$; then, carry the result to the next time step $t_{i+1}$: $\Lparam{\OT}{t_{i+1}}{0} \gets \Lparam{\OT}{t_i}{K}$.
This procedure is repeated until the last time step $t_T$.

The intuition behind this construction is that the obtained parameter $\Lparam{\OT}{t_T}{K}$ is ``close'' to previous steps along the time sequence.
Since the effect of catastrophic forgetting in neural networks can be mitigated with gradient trajectory regularization~\citep{goodfellowEmpiricalInvestigationCatastrophic2015,chaudhryRiemannianWalkIncremental2018}, the shortest gradient path helps to retain useful information from previous time steps.

\subsubsection{Meta-optimization}
LeapRec combines GTL and OTL by taking the summation of their prediction scores:
\begin{equation}
    \label{eq:meta-pred-score}
    \func{f}_{\{\GT,\OT\}} (u, i) = \func{f}^{\textsf{GTL}}_\GT (u, i) + \func{f}^{\textsf{OTL}}_\OT (u, i).
\end{equation}
To train $\{\GM, \OM\}$, we simultaneously update $\{\GT, \OT\}$ and update $\{\GM, \OM \}$ at the end of an iteration using Leap, as shown in line 13 of Algorithm~\ref{alg:training}. In deployment, we use the latest model parameters $\{\Lparam{\GT}{t_\latestTime}{K},\Lparam{\OT}{t_\latestTime}{K}\}$ for making recommendations.

\begin{figure}[t]
    \centering
    \includegraphics[width=.7\linewidth]{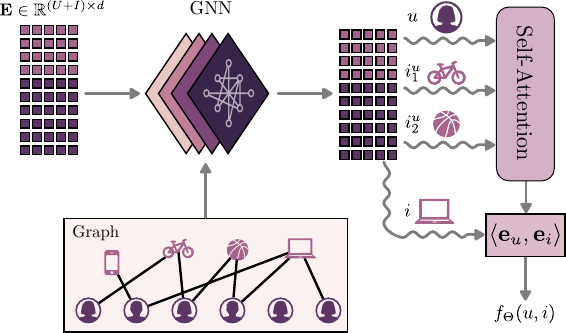}
    \caption{Our neural recommender. $\Theta$ denotes a set of embedding $\emb{E}$, GNN, and Self-Attention parameters.}
    \label{fig:recommender}
\end{figure}

\subsection{Neural Recommender}

To demonstrate the application of LeapRec to a recommender system, we define a neural recommender described in Figure~\ref{fig:recommender}.
The architecture consists of two main components: a graph neural network (GNN)~\citep{scarselliGraphNeuralNetwork2009,kipfSemiSupervisedClassificationGraph2016} and a self-attention mechanism (SA)~\citep{vaswaniAttentionAllYou2017}.

The neural recommender learns an embedding matrix $\emb{E} \in \mathbb{R}^{(U+I) \times d}$, where each row corresponds to a $d$-dimensional embedding of a user ($\mathbf{e}_u \in \mathbb{R}^d$) or an item ($\mathbf{e}_i \in \mathbb{R}^d$). Note that GTL and OTL can have different numbers of dimensionalities denoted by $d^{\textsf{GTL}}$ and $d^{\textsf{OTL}}$, respectively.
The GNN component refines initial features of users and items based on their related information guided by an interaction graph.
On the other hand, the SA component extracts information from the histories of users to enhance user embeddings for more accurate recommendations.

\newcommand{\nb}[1]{\mathcal{N}_{#1}}
\subsubsection{Graph neural network}
Let $\Graph = (\Node, \Edge)$ be a graph of user-item interactions.
The node set $\Node$ consists of all users and items in data $\data$, and the edge set $\Edge$ consists of interactions between users and items.
Then, we use the GNN to propagate the node information based on the graph structure.
The propagation for any node, namely $m$, in the graph by the $\ell^{th}$ layer of the GNN can be written as:
\begin{align}
    \emb{e}^{\ell}_m \gets \sigma ( a^{\ell}_{m,m} \emb{W}^{\ell}_1 \emb{e}^{\ell-1}_m + \sum_{n \in \nb{m}} a^{\ell}_{m,n} \emb{W}^{\ell}_2 \emb{e}^{\ell-1}_n ),
\end{align}
where $\nb{m}$ denotes a set of the neighborhood of node $m$, $\sigma$ denotes a non-linearity function, $\emb{W}^{\ell}_1, \emb{W}^{\ell}_2$ are transformation matrices of the $\ell^{th}$ layer, and $a^{\ell}_{m,n}$ is a normalizing factor; for example, GCN~\citep{kipfSemiSupervisedClassificationGraph2016} uses $a^{*}_{m,n} = 1 / {\sqrt{\num{\nb{m}} \num{\nb{n} }}}$; 
or GAT~\citep{velickovicGraphAttentionNetworks2018} proposed to use an attention mechanism to train $a^\ast_{m,n}$.
The final outputs are stored in the embedding matrix $\emb{E}$.

\subsubsection{Self-Attention} 
We construct the sequence of items that a user, namely $u$, has been interacted with before timestamps $t$: $\Seq^u = (i^u_1, i^u_2, \ldots, i^u_\num{\Seq^u})$.
Embedding lookup is performed on $E$ for each item in the sequence: $\emb{x}^u_j = \emb{e}_{i^u_j}$.
We also append the embedding of user $u$ to the sequence: $\emb{x}^u_{\num{\Seq^u}+1} = \emb{e}_u$.
Then, we use SA to fuse the information inside the sequence.
The attention score between inputs, namely from $\emb{x}^u_k$ to $\emb{x}^u_j$, and the output is calculated as
\begin{align*}
    {a}^u_{k \to j} &= \frac{(\emb{W}^Q \emb{x}^u_j)^{\intercal}(\emb{W}^K \emb{x}^u_k)}{\sqrt{d}}, \, \tilde{{a}}^u_{k \to j} = \frac{ {a}^u_{k \to j} }{ \sum_{k'=1}^{\num{\Seq^u}+1} {a}^u_{k' \to j} }, \\[-10pt]
    \emb{z}^u_j &= \sum_{k=1}^{\num{\Seq^u}+1} \tilde{{a}}^u_{k \to j} \emb{W}^V  \emb{x}^u_k ,
\end{align*}
where $\emb{W}^Q, \emb{W}^K, \emb{W}^V \in \R^{d \times d} $ are the transformation matrices.
Note that our SA cannot be aware of the positions in a sequence\footnote{The purpose is to highlight the contribution of LeapRec. We provide a comparison with the neural recommender ``with position-aware SA'' and ``without LeapRec'' variants in the Appendix.}, unlike those SA-based methods for the sequential recommendation.
After each SA layer, we apply each feature vector to a two-layer fully-connected feed-forward network (FFN) with ReLU activation.
For each SA and FFN, we use a residual connection, dropout, and layer normalization, following~\citet{vaswaniAttentionAllYou2017}:
\begin{equation*}
    \emb{x}^u_j \gets \text{LayerNorm} ( \text{Dropout} (\emb{z}^u_j) + \emb{x}^u_j).
\end{equation*}

\subsubsection{Prediction}
Thus far, we have defined the components of the neural recommender (Figure~\ref{fig:recommender}).
To find the compatibility score between a user $u$ and an item $i$, 
we obtain the embedding $\mathbf{e}_u$ from the last element of the sequence $\emb{x}^u_{\num{S^u}+1}$,
and embedding $\mathbf{e}_i$ from the output of the GNN.
The score between $u$ and $i$ is given by the dot product  $\langle \emb{e}_u, \emb{e}_i \rangle_{\mathbb{R}^d}$.
This is a concrete modeling for $f_{\OT}^\textsf{OTL}$ and  $f_{\GT}^\textsf{GTL}$ in Eq.~\ref{eq:meta-pred-score}.

Our neural recommender with LeapRec is a powerful model that unifies multiple concepts of recommender systems;
GNN signifies the similarity between users and items beyond local interaction structures, which helps in learning common patterns; SA captures dependencies between items; LeapRec enhances both aforementioned to learn temporal dynamics. This is achievable because of the minimal computation cost of LeapRec.

\section{Experiments}
\label{sec:experiments}

We conducted experiments to answer the following research questions:
\textbf{RQ1} How well does LeapRec perform compared with SOTA models?
\textbf{RQ2} How effective is LeapRec as a plug-ins method for GNN and SA models?
\textbf{RQ3} How important are GTL and OTL, and what are the changes in their learned representations over time?
\textbf{RQ4} How sensitive is LeapRec to hyperparameters, and can LeapRec converge?
\subsection*{Experimental Settings}

\subsubsection*{Dataset}
\begin{figure}[tb]
    \centering
    \begin{subfigure}{.3\linewidth}
        \includegraphics[width=\linewidth]{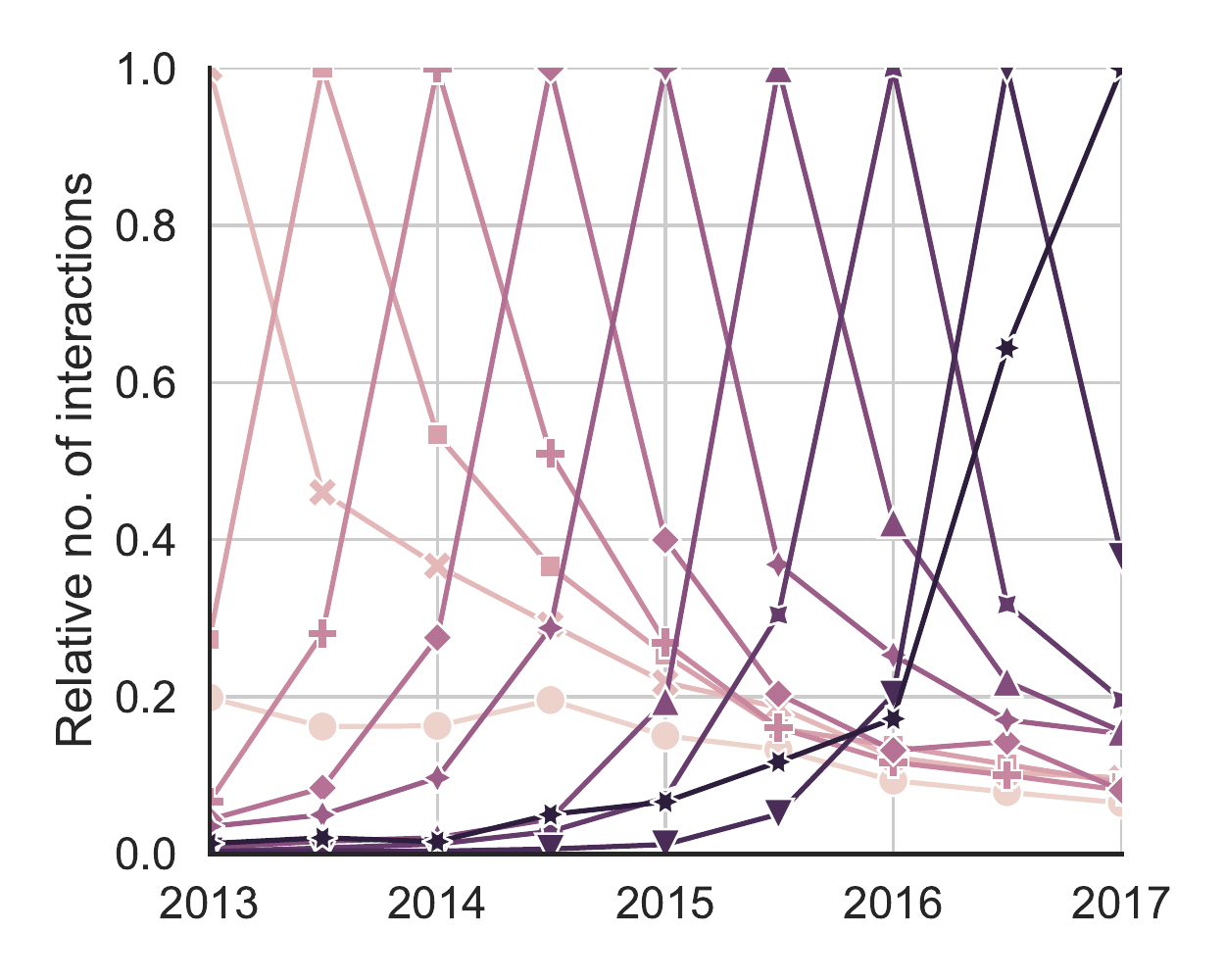}
        \caption{Amazon}
        \label{fig:dataset:top-interactions:amz}
    \end{subfigure}
    \begin{subfigure}{.3\linewidth}
        \includegraphics[width=\linewidth]{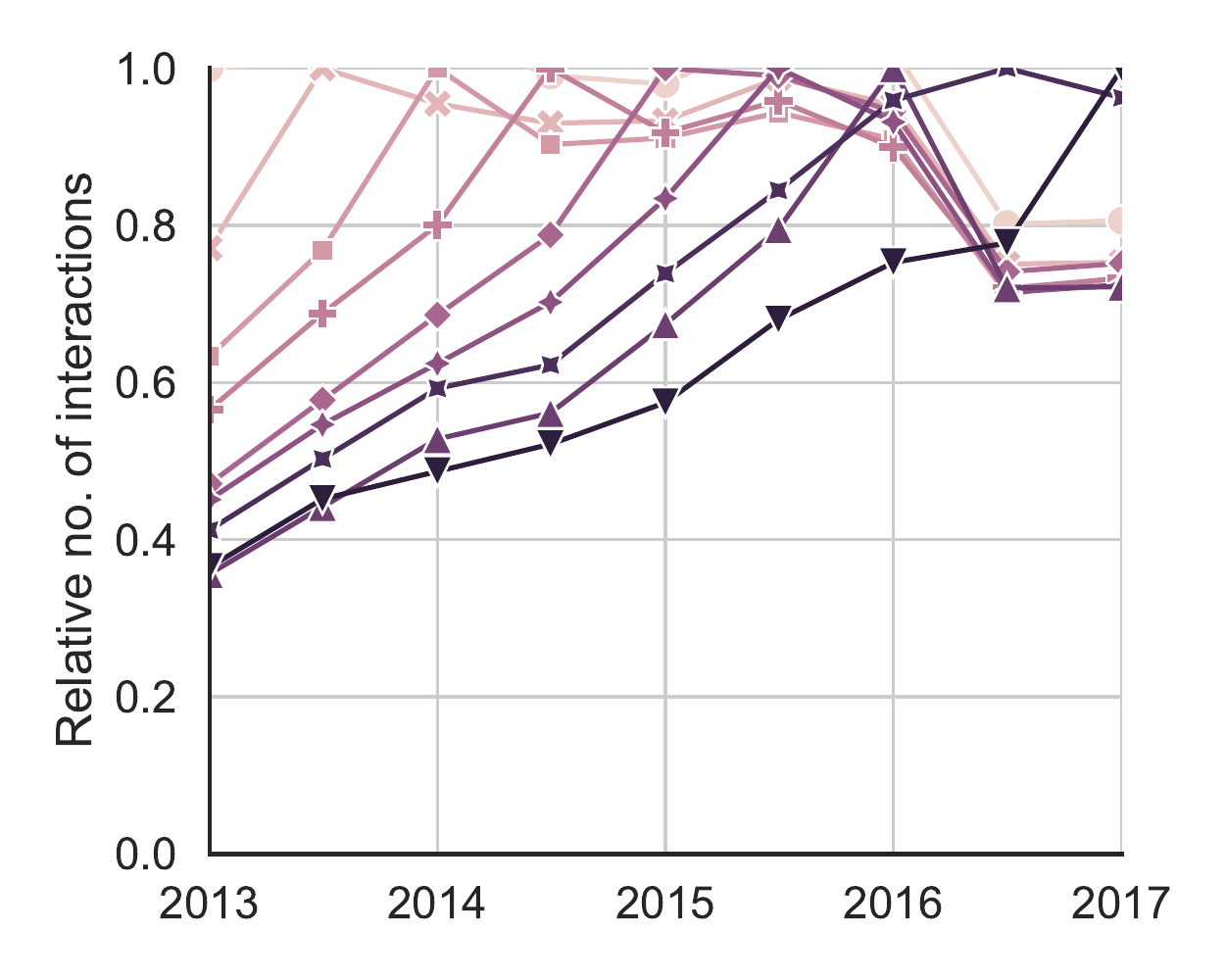}
        \caption{Yelp}
        \label{fig:dataset:top-interactions:yelp}
    \end{subfigure}
    \caption{Relative number of interactions of groups of popular items (top 100) under each timestamp. Each line represents a group of items that became popular at the same time. The popularity of items on Amazon changed more drastically than one on Yelp.}
    \label{fig:dataset:top-interactions}
\end{figure}

We used three publicly available datasets: Amazon~\citep{niJustifyingRecommendationsUsing2019}, Goodreads~\citep{wangRecurrentRecommendationLocal2019}, and Yelp\footnote{\url{https://www.yelp.com/dataset}, updated February 2021.}. 
User-item interactions in the Amazon dataset are product reviews.
Our experiments used the preprocessed Amazon data by \citet{wangNextitemRecommendationSequential2020}, where data across categories are mixed.
The Goodreads dataset is mined from the book-reading social network Goodreads.
User-item interactions here are ratings and reviews of books.
Yelp is a crowd-sourced business reviewing website. 
We used the latest officially provided data, updated on February 2021.
For each dataset, we set a cutting time that only the data occurred before the time was used for training and the rest for validation/testing.
The basic data statictics are presented in Table~\ref{tab:ds-stat}.
To visualize the difference in the temporal dynamics between datasets, we plot the relative number of interactions in each group representing the top 100 items grouped by \emph{peak} time on Amazon and Yelp.
Figure~\ref{fig:dataset:top-interactions} shows that popular items on Amazon changed more frequently compared with those on Yelp.
\begin{table}[tb]
    \centering
    \small
    \begin{tabular}{ccccc}
        \toprule
        Dataset & \# Users & \# Items & \# Actions & {Cutting time} \\
        \midrule
        Amazon & 70,097 & 56,708 & 1,540,911 & 2017/06 \\
        Goodreads & 16,862 & 20,633 & 1,762,094 & 2015/12 \\
        Yelp & 234,664 & 45,032 & 3,880,015 & 2018/12 \\
        \bottomrule
    \end{tabular}
    \caption{Basic statistics of datasets. Cutting times determine train/test splits.}
    \label{tab:ds-stat}
\end{table}

\newcommand{\static}{{\scriptsize \faStop}}
\newcommand{\sequence}{{\scriptsize \faForward}}
\newcommand{\temporal}{{\scriptsize \faClock[regular]}}
\newcommand{\stream}{{\scriptsize \faPlayCircle}}
\begin{table*}[t]
    \centering
    \small
      \begin{adjustbox}{max width=\textwidth}
      \begin{tabular}{lcccc cccccccc}
      \toprule
      \multirow{2}[4]{*}{Model} & \multicolumn{4}{c}{Amazon} & \multicolumn{4}{c}{Goodreads} & \multicolumn{4}{c}{Yelp} \\
    
      \cmidrule(lr){2-5} \cmidrule(lr){6-9} \cmidrule(lr){10-13}      & \scriptsize{HR@1}  & \scriptsize{HR@5}  & \scriptsize{NDCG@5}  & \scriptsize{MRR} & \scriptsize{HR@1}  & \scriptsize{HR@5}  & \scriptsize{NDCG@5}  & \scriptsize{MRR} & \scriptsize{HR@1}  & \scriptsize{HR@5}  & \scriptsize{NDCG@5}  & \scriptsize{MRR} \\
      \midrule
    \static{} MF & 0.0820 & 0.2413 & 0.1629 & 0.1756 & 0.2008 & 0.5215 & 0.3662 & 0.3525 & 0.2847 & 0.6637 & 0.4823 & 0.4544 \\
    \static{} NCF & 0.0751 & 0.2239 & 0.1506 & 0.1638 & 0.2003 & 0.5158 & 0.3630 & 0.3498 & 0.2923 & 0.6620 & 0.4857 & 0.4558 \\
    \static{} BiVAE & 0.0891 & 0.2587 & 0.1748 & 0.1868 & 0.1966 & 0.5021 & 0.3542 & 0.3429 & 0.3056 & 0.6672 & 0.4954 & 0.4678 \\
    \static{} NGCF & 0.0751 & 0.2273 & 0.1523 & 0.1646 & 0.1978 & 0.5146 & 0.3612 & 0.3482 & 0.2794 & 0.6545 & 0.4751 & 0.4463 \\
    \static{} LightGCN & 0.0883 & 0.2517 & 0.1713 & 0.1826 & 0.2114 & 0.5342 & 0.3785 & 0.3632 & \underline{0.3254} & 0.6930 & 0.5186 & 0.4875 \\
    \sequence{} GRU4Rec & 0.0851 & 0.2681 & 0.1777 & 0.1890 & 0.2087 & 0.5194 & 0.3694 & 0.3569 & 0.3159 & 0.7145 & 0.5252 & 0.4895 \\
    \sequence{} NARM & 0.0890 & 0.2764 & 0.1838 & 0.1945 & 0.2171 & 0.5340 & 0.3811 & 0.3669 & 0.3124 & 0.7160 & 0.5239 & 0.4879 \\
    \sequence{} SUM & 0.0938 & 0.2896 & 0.1928 & 0.2025 & 0.2070 & 0.5224 & 0.3697 & 0.3563 & 0.2858 & 0.6906 & 0.4963 & 0.4627 \\
    \sequence{} SASRec & 0.0879 & 0.2714 & 0.1806 & 0.1917 & 0.2147 & 0.5254 & 0.3756 & 0.3627 & 0.3181 & \underline{0.7215} & \underline{0.5295} & \underline{0.4932} \\
    \sequence{} MetaTL** & 0.0748 & 0.2267 & 0.1515 & 0.1652 & 0.1930 & 0.4989 & 0.3506 & 0.3397 & 0.2849 & 0.6704 & 0.4860 & 0.4560 \\
    \temporal{} TGN & 0.1025 & \underline{0.3227} & \underline{0.2140} & \underline{0.2200} & 0.1623 & 0.4448 & 0.3077 & 0.3021 & 0.2486 & 0.6536 & 0.4576 & 0.4278 \\
    \temporal{} SLi-Rec & 0.0980 & 0.3005 & 0.2007 & 0.2097 & \underline{0.2264} & \underline{0.5509} & \underline{0.3947} & \underline{0.3785} & 0.2999 & 0.7103 & 0.5140 & 0.4781 \\
    \temporal{} TiSASRec & 0.0888 & 0.2755 & 0.1832 & 0.1942 & 0.2176 & 0.5337 & 0.3814 & 0.3674 & 0.3117 & 0.7106 & 0.5205 & 0.4850 \\
    \temporal{} HyperRec & \underline{0.1056} & 0.3023 & 0.2057 & 0.2143 & 0.2143 & 0.5282 & 0.3769 & 0.3633 & 0.2793 & 0.6968 & 0.4966 & 0.4608 \\
    \stream{} SPMF & 0.0587 & 0.1867 & 0.1232 & 0.1364 & 0.1366 & 0.3402 & 0.2415 & 0.2430 & 0.1526 & 0.3851 & 0.2731 & 0.2681 \\
    \stream{} SML* & 0.1029 & 0.3031 & 0.2045 & 0.2115 & 0.1892 & 0.4776 & 0.3377 & 0.3294 & 0.2211 & 0.5533 & 0.3923 & 0.3785 \\
    \rule{0pt}{2ex}    
    \textbf{LeapRec} & \textbf{0.1252} & \textbf{0.3484} & \textbf{0.2391} & \textbf{0.2443} & \textbf{0.2452} & \textbf{0.5760} & \textbf{0.4171} & \textbf{0.3991} & \textbf{0.3420} & \textbf{0.7466} & \textbf{0.5547} & \textbf{0.5165} \\
       \cmidrule(lr){2-5} \cmidrule(lr){6-9} \cmidrule(lr){10-13} 
      \emph{Improvement} & 18.59\% & 7.97\% & 11.74\% & 11.07\% & 8.30\% & 4.55\% & 5.67\% & 5.44\% & 5.09\% & 3.48\% & 4.76\% & 4.72\% \\
      \bottomrule
      \end{tabular}%
    \end{adjustbox}
      
    \caption{Performance comparison of different models. The best performances are highlighted in bold. The second best performances are underlined. \emph{Improvement} indicates \emph{relative} improvement of LeapRec over the best performing baseline in each case. The symbols \static, \sequence, \temporal, and \stream\ denote static, sequential, temporal, and streaming models, respectively. We mark MAML approaches by ** for vanilla MAML and * for first-order MAML (FOMAML).}      
    \label{tab:exp-sota}%
  \end{table*}%

\subsubsection*{Evaluation protocol}
We divided each dataset into training, validation, and test sets based on the timestamps of interactions.
We kept the interactions within six months after the cutting time as the validation set, and the rest after that as the test set.
For each interaction in the test set, we randomly sampled 99 negatives instead of ranking all items to reduce the computation cost in the evaluation, and subsequently, the positive will be ranked among them by a model.
We measured the recommendation quality using commonly used metrics for Top-K recommendation including Hit Rate (HR@K), Normalized Discounted Cumulative Gain (NDCG@K), and Mean Reciprocal Rank (MRR).

\subsection*{Performance Comparison (RQ1)}
To answer \textbf{RQ1}, we extensively evaluated LeapRec by comparing it to various types of recommendation models:
\begin{itemize}
    \item[\static] \textbf{Static models} are unaware of the temporal change in the behaviors of users. This category includes matrix factorization (MF)~\citep{korenMatrixFactorizationTechniques2009}, deep neural network-based: NCF~\citep{heNeuralCollaborativeFiltering2017}, variational autoencoder-based: BiVAE~\citep{truongBilateralVariationalAutoencoder2021}, and GNN-based: NGCF~\citep{wangNeuralGraphCollaborative2019} and LightGCN~\cite{heLightGCNSimplifyingPowering2020} methods. 
    \item[\sequence] \textbf{Sequential models} are partially aware of the dynamic in the user behaviors based on the change in input sequences, this includes RNN-based approaches: GRU4Rec~\citep{hidasiSessionbasedRecommendationsRecurrent2016} and NARM~\citep{liNeuralAttentiveSessionbased2017}, a Memory-Network-based approach: SUM~\citep{lianMultiInterestAwareUserModeling2021a}, a SA-based model: SASRec~\citep{kangSelfAttentiveSequentialRecommendation2018},
    and a meta-learning-based method: MetaTL~\citep{wangSequentialRecommendationColdstart2021}.
    \item[\temporal] \textbf{Temporal models} are explicitly aware of temporal dynamics or include temporal data in learning representations. This includes GNN-based: TGN~\citep{tgn_icml_grl2020}, RNN-based: SLi-Rec~\citep{yuAdaptiveUserModeling2019}, SA-based: TiSASRec~\citep{liTimeIntervalAware2020} and GNN with SA-based: HyperRec~\citep{wangNextitemRecommendationSequential2020}
    \item[\stream] \textbf{Streaming models} inevitably favor recent data owing to the limited data access. We presents the results for SPMF~\citep{wangStreamingRankingBased2018} and SML~\citep{zhangHowRetrainRecommender2020}. Note that SML uses FOMAML to train a meta-updater.
\end{itemize}

\subsubsection*{Hyper-parameter settings} For all baselines, we conducted hyperparameter search on the dimensionality of embedding: $\{ 64,128 \}$, learning rate: $\{0.001, 0.0001\}$, and dropout: $\{ 0, 0.2, 0.5\}$.
We also run hyperparameter search on the number of layers $L$: $\{1,2,3,4\}$ for GNN and $\{1,2\}$ for SA.
We used the default values as provided for the rest.
We used Adam~\citep{kingma2014adam} as the optimizer for all models.
For LeapRec, we used the same hyperparameter settings as SASRec for its SA, and for its GNN, we adopt two-layer GCN~\cite{kipfSemiSupervisedClassificationGraph2016} with dropout rate set at 0.2.
We performed grid search on meta learning rates, the number of update steps $K$, and the time granularity on Amazon, then used them on other datasets\footnote{After the grid search, we set learning rates: $\beta=\eta=0.01$. For the best performance to answer \textbf{RQ1}, we set $K$ to 40 and time granularity to one month, whereas for faster training to answer \textbf{RQ2-4}, we set $K$ to 20 and the time granularity to two months.}.
The grid search for dimensionalities of OTL and GTL was performed for each dataset with values $\{0, 40, 91, 121, 128\}$ (see Table~\ref{tab:exp-ndim}).

\subsubsection*{Experimental results}
We report the experimental results in Table~\ref{tab:exp-sota}.
\textbf{LeapRec consistently outperforms all the baselines across all datasets and metrics.}
Specifically, LeapRec achieves 8\% to 18\% relative improvements over the best performing baseline on Amazon dataset.
This indicates that LeapRec can adapt fast enough to capture the dynamics between timestamps.
Similarly, we observe that temporal models tend to outperform static and sequential models on the Amazon dataset. 
Conversely, on Yelp dataset, they achieve lower performances than those of sequential models.
This observation demonstrates the inflexibility of the current temporal-aware SOTA recommendation models in modeling temporal dynamics.
However, LeapRec performs well in both all datasets, indicating its flexibility.

\subsection*{LeapRec as a Plug-ins Method (RQ2)}
Recently, GNN- and SA-based models have gained growing attention from researchers.
It is interesting to determine the effect of LeapRec on these neural networks as a plug-ins method.
In this experiment, we show that LeapRec can be applied to various types of models, making them temporal-aware.
In the GNN-based category, we applied LeapRec on GCN, NGCF, and LightGCN, and compare the recommendation results between these GNNs with and without \model.
The main difference between these GNNs is based on feature transformation. LightGCN, a SOTA GNN for recommendation, is a simple GNN without feature transformation and nonlinearity.
We also investigate the importance of Leap by introducing a FOMAML variant based on the modification of Eq.~\ref{eq:leap:meta_grad}.
\begin{figure}[tb]
    \centering
    \includegraphics[width=.7\linewidth]{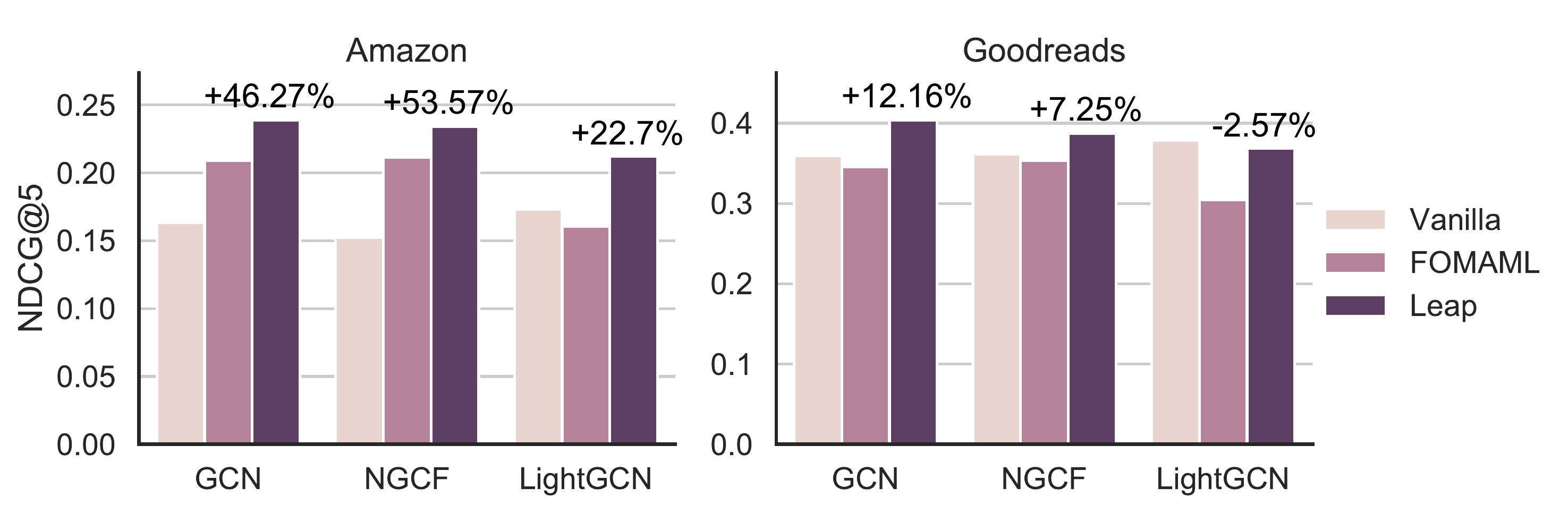}
    \caption{Performance of GNNs. FOMAML and Leap variants were equipped with GTL and OTL. The number over a bar indicates the relative improvement over its vanilla.
    }
    \label{fig:result:gnn}
\end{figure}

We plot the results of the GNN experiments in Figure~\ref{fig:result:gnn}.
The NDCG@5 scores significantly increase up to 46.27\% and 53.57\% in relative gain among GCN and NGCF thanks to LeapRec.
For LightGCN, we observe an improvement on the Amazon dataset and a slight decrease on Goodreads.
We suppose that the feature transformation can accelerate the task-learning process~\citep{leeGradientBasedMetaLearningLearned2018,flennerhagMetaLearningWarpedGradient2019}.
In addition, \textbf{the LeapRec variants significantly outperform the FOMAML variants in all cases.}
This empirically confirms the importance of trajectory-based meta-learning in the recommendation context.

\begin{table}[tb]
    \centering
    \small
      \begin{tabular}{lcccc}
      \toprule
      \multirow{2}[4]{*}{Model} & \multicolumn{2}{c}{Amazon} & \multicolumn{2}{c}{Goodreads } \\
     \cmidrule(lr){2-3} \cmidrule(lr){4-5}     & $L$=1 & $L$=2 & $L$=1 & $L$=2 \\
      \midrule
      SASRec & 0.1702 & 0.1806 & 0.3701 & 0.3709 \\
      TiSASRec & 0.1774 & 0.1832 & 0.3709 & 0.3814 \\
      \model-SASRec & 0.1992 & \textbf{0.2045} & \textbf{0.3956} & 0.3934 \\
     \cmidrule(lr){2-3} \cmidrule(lr){4-5}
      \emph{Improvement} & 12.26\% & 11.63\% & 6.66\% & 3.15\% \\
      \bottomrule
      \end{tabular}%
    \caption{Performance (NDCG@5) of SASRec without and with \model{} with different numbers of layers $L$}
    \label{tab:exp-sas}%
  \end{table}%

The results of experiments on SA-based models are reported in Table~\ref{tab:exp-sas}.
Even though SA models can already adapt to recent timestamps, \model{} still improves SA with up to 12.26\% gain over temporal-aware TiSASRec.
This indicates the benefit of LeapRec's adaptability. 
We observe similar improvements to GNN and SA models on the Yelp dataset (omitted due to the page limit).

\subsection*{Analysis on Global and Ordered Time Leap (RQ3)}
We investigate the effects of GTL and OTL on the performance of the model and the best resource-allocation ratio between them. 
We run experiments of LeapRec with varying the dimensionalities of GTL and OTL while maintaining the floating-point operations per second (FLOPs) of the model.
The results are reported in Table~\ref{tab:exp-ndim}. We observe that the best configuration depends on the datasets. 
In fast dynamic data such as Amazon, the model can perform well without GTL. 
Conversely, in slower dynamic data (Goodreads, Yelp), GTL becomes more important.

\begin{table}[t]
    \centering
    \small
      \begin{tabular}{ccccc}
      \toprule
      \multicolumn{2}{c}{Dimensionality ($d$)} & \multicolumn{3}{c}{Dataset} \\
      \cmidrule(lr){1-2} \cmidrule(lr){3-5}
      GTL ($d^{\textsf{GTL}}$) & OTL ($d^{\textsf{OTL}}$) & Amazon & Goodreads  & Yelp  \\
      \midrule
      128 & unused & 0.2220 & 0.3885 & 0.5238 \\
      121 & 40 & 0.2339 & 0.4061 & \textbf{0.5485} \\
      91 & 91 & 0.2384 & 0.4025 & 0.5433 \\
      40 & 121 & 0.2415 & \textbf{0.4093} & 0.5366 \\
      unused & 128 & \textbf{0.2416} & 0.3974 & 0.5379 \\
      \bottomrule
      \end{tabular}%
    \caption{LeapRec with different combinations of GTL and OTL dimensionalities with similar FLOPs in total.}
    \label{tab:exp-ndim}%
  \end{table}%
 % This experiment supports the claim of Theorem 3. 

\begin{figure}[tb]
    \centering
    \includegraphics[width=.6\linewidth]{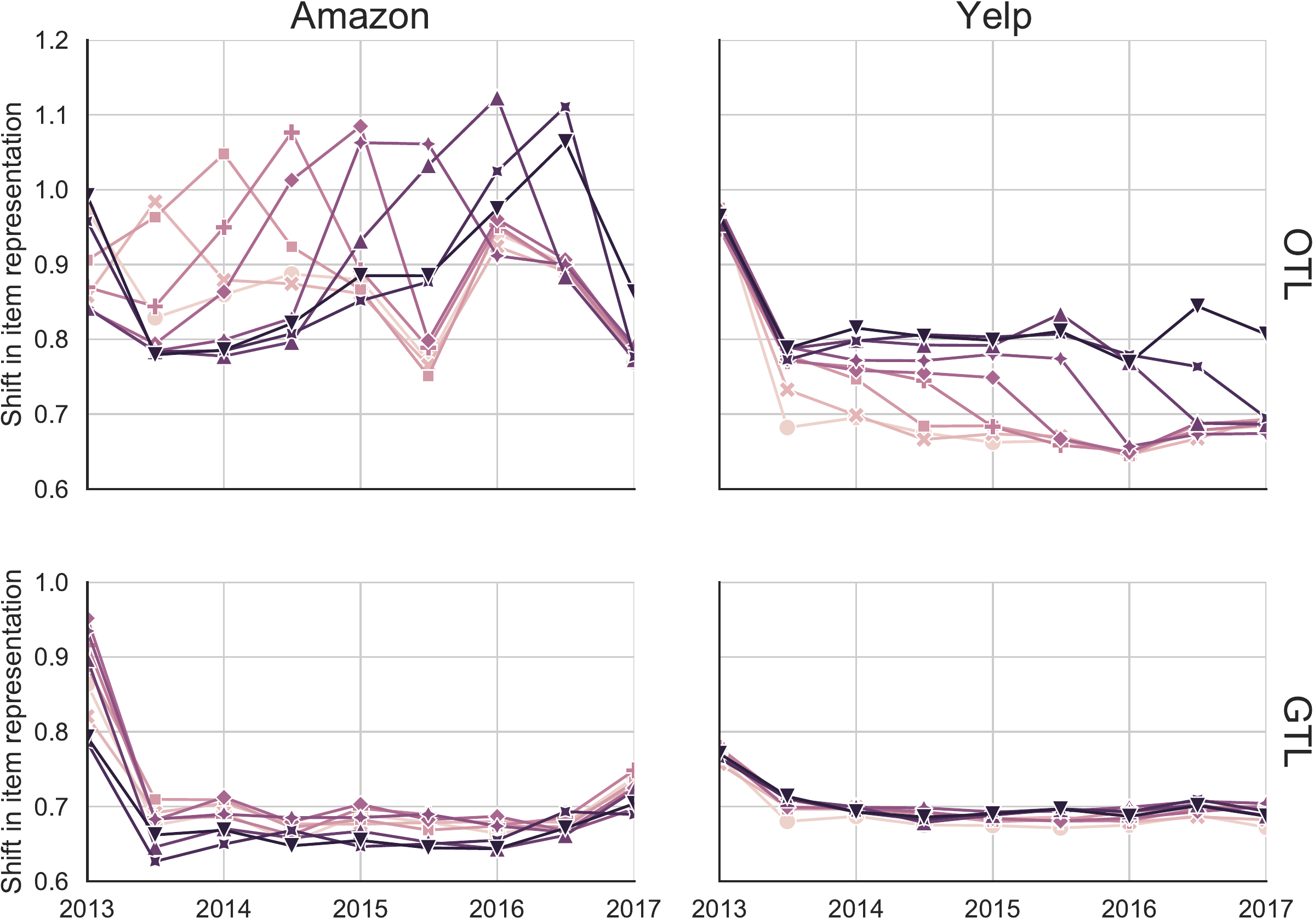}
    \caption{Shift in the item representations monitored on OTL and GTL of popular items under each timestamp. Each line represents a group of items that became popular at the same time (corresponding to the same group in Figure~\ref{fig:dataset:top-interactions}).}
    \label{fig:embedding}
\end{figure}

To further investigate the difference, we monitored a shift in the final representation of popular items in both GTL and OTL, as motivated by the observation in Figure~\ref{fig:dataset:top-interactions}. The shift of the representation of item $i$ from time $t_{j-1}$ to $t_{j}$ is defined as:
   $s_{t_{j}}^i = \norm{{\hat{\emb{e}}_{t_{j},i}^K} -  {\hat{\emb{e}}_{t_{j-1},i}^K}}$, where $\hat{\emb{e}}$ denotes the normalized vector of $\emb{e}$. Remind that $K$ is the number of update steps.

We plot the shift monitored on OTL and GTL in Figure~\ref{fig:embedding}.
For OTL, we observe that the popular group at a certain time had a larger shift than less popular groups. This implies that OTL \emph{actively} adjusted such a group more than others to adapt to the trends shown in Figure~\ref{fig:dataset:top-interactions}. As time went on, the popularity of such a group would decrease or become stable, and OTL also became less active accordingly. In addition, we observe that OTL was more active on Amazon than Yelp because of the difference in the magnitude of temporal dynamics. On the other hand, for GTL, its activeness was almost constant across time. These demonstrate that GTL holds common and long-term patterns while OTL captures temporal information that varies frequently. 

\subsection*{Sensitivity and Convergence Analysis (RQ4)}

\begin{figure}[tb]
    \centering
        \includegraphics[width=.3\linewidth]{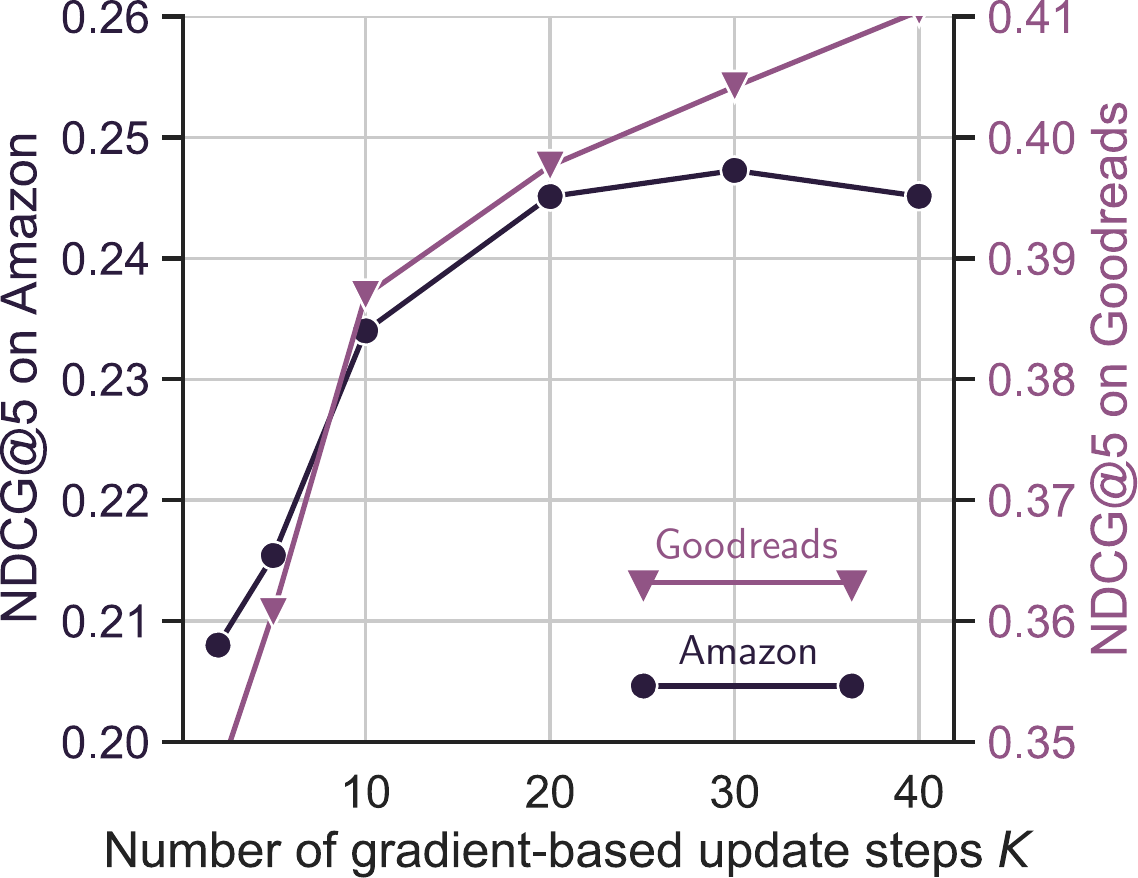}
        \hspace{0.05\textwidth}
        \includegraphics[width=.3\linewidth]{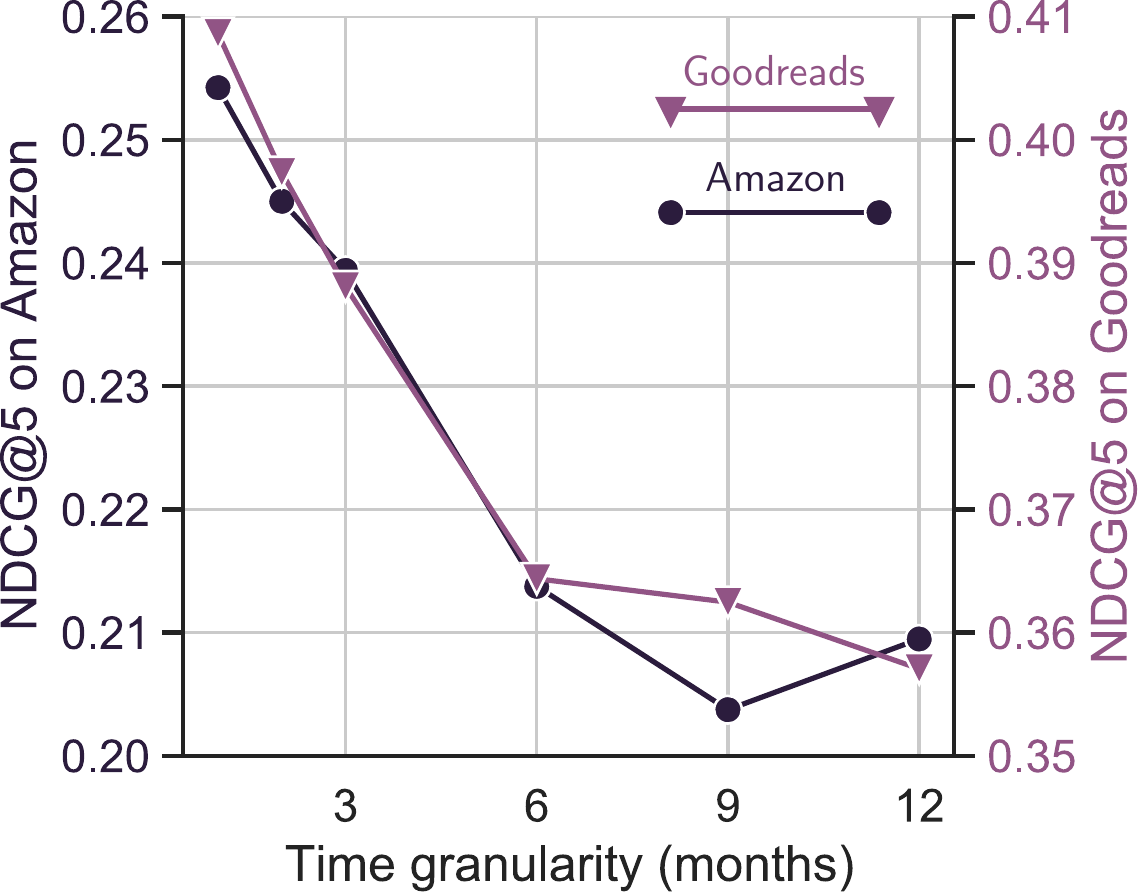}
    \caption{Hyperparameter sensitivity analyses}
    \label{fig:hyper}
\end{figure}

We investigate the effect of the number of gradient-based update steps $K$ and time granularity of timestamps on the performance of LeapRec. 
We conducted further experiments using LeapRec with varying numbers of both and plot the results in Figure~\ref{fig:hyper}. 
We observe that the performance of LeapRec tends to increase with the number of update steps. 
The performance on Goodreads continued to increase when $K>20$, but reached a plateau on Amazon. 
This possibly can be contributed to the difference in the densities of the datasets.
With respect to time granularity, LeapRec performs best with the finest-grained of one month. 
This implies its fast adaptability.
Unsurprisingly, LeapRec behaves closer to a static model with coarser-grained time granularity.

\begin{figure}[tb]
    \centering
    \includegraphics[width=.7\linewidth]{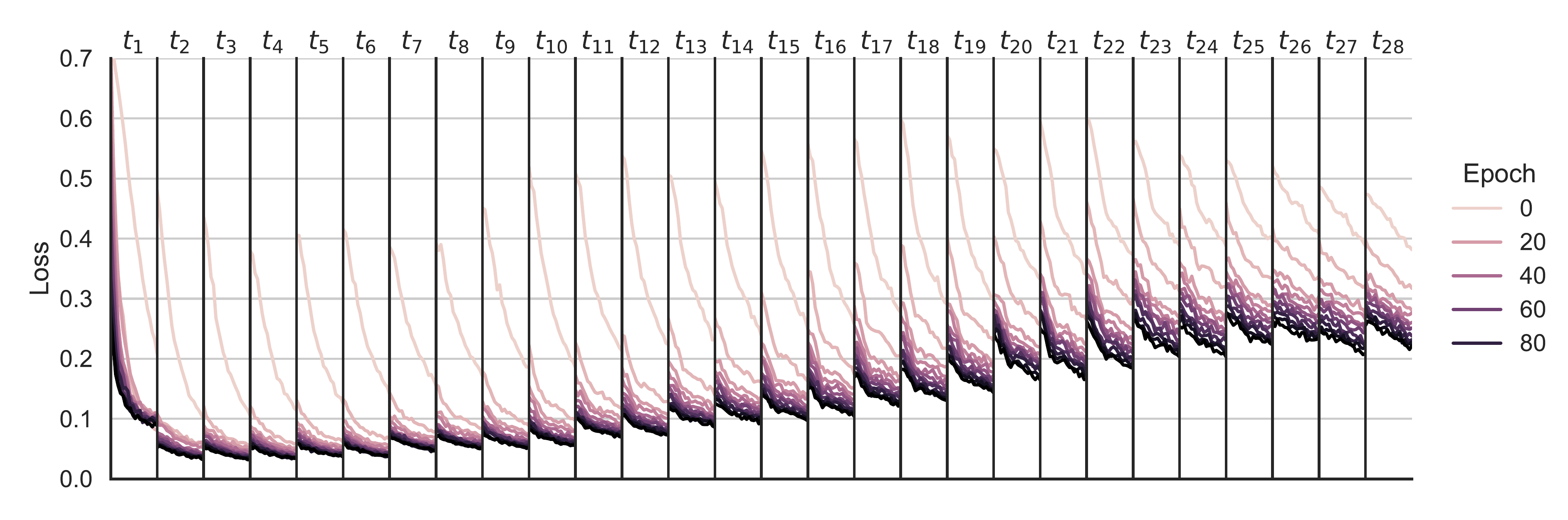}
    \caption{Training loss of LeapRec in each timestamp on Amazon dataset in different epochs. The horizontal axis in each time window is the gradient-based update steps.}
    \label{fig:training:both}
\end{figure}

Finally, we analyze the convergence of LeapRec by monitoring its training loss. We plot the training loss across timestamps in Figure~\ref{fig:training:both}. We observe that the training loss continued to decrease on all timestamps, suggesting that LeapRec could converge.

\section{Conclusion}

In this work, we proposed a novel method using trajectory-based meta-learning to solve the temporal dynamic challenge in modern recommender systems.
The empirical results clearly show the advantages of our method compared with a wide range of SOTA methods.
Our deep analyses into the temporal dynamics of benchmark datasets also validate our designs of GTL and OTL as they capture long-term and short-term information respectively.
We believe the effectiveness of our proposal will initiate new discussions regarding trajectory-based recommendation.

\section*{Acknowledgments}
We would like to thank Noppayut Sriwatanasakdi and Nontawat Charoenphakdee for helpful discussion.
This work was partially supported by JST CREST (Grant Number JPMJCR1687), JSPS Grant-in-Aid for Scientific Research (Grant Number 17H01785, 21K12042), and the New Energy and Industrial Technology Development Organization (Grant Number JPNP20006).
Nuttapong Chairatanakul was supported by MEXT scholarship.

\bibliographystyle{aaai22}
\bibliography{aaai22auto}

\begin{thebibliography}{64}
\providecommand{\natexlab}[1]{#1}

\bibitem[{Aggarwal(2016)}]{Aggarwal2016}
Aggarwal, C.~C. 2016.
\newblock Model-Based Collaborative Filtering.
\newblock In \emph{Recommender Systems: {{The}} Textbook}, 71--138. {Springer
  International Publishing}.

\bibitem[{Ahlberg, Nilson, and Walsh(1967)}]{ahlberg1967theory}
Ahlberg, J.~H.; Nilson, E.~N.; and Walsh, J.~L. 1967.
\newblock The theory of splines and their applications.
\newblock \emph{Mathematics in science and engineering}.

\bibitem[{Chairatanakul, Murata, and
  Liu(2019)}]{chairatanakulRecurrentTranslationBasedNetwork2019}
Chairatanakul, N.; Murata, T.; and Liu, X. 2019.
\newblock Recurrent {{Translation}}-{{Based Network}} for {{Top}}-{{N Sparse
  Sequential Recommendation}}.
\newblock \emph{IEEE Access}, 7: 131567--131576.

\bibitem[{Chang et~al.(2017)Chang, Zhang, Tang, Yin, Chang, {Hasegawa-Johnson},
  and Huang}]{changStreamingRecommenderSystems2017}
Chang, S.; Zhang, Y.; Tang, J.; Yin, D.; Chang, Y.; {Hasegawa-Johnson}, M.~A.;
  and Huang, T.~S. 2017.
\newblock Streaming {{Recommender Systems}}.
\newblock In \emph{{{WWW}} '17}, 381--389.

\bibitem[{Chaudhry et~al.(2018)Chaudhry, Dokania, Ajanthan, and
  Torr}]{chaudhryRiemannianWalkIncremental2018}
Chaudhry, A.; Dokania, P.~K.; Ajanthan, T.; and Torr, P. H.~S. 2018.
\newblock Riemannian {{Walk}} for {{Incremental Learning}}: {{Understanding
  Forgetting}} and {{Intransigence}}.
\newblock In Ferrari, V.; Hebert, M.; Sminchisescu, C.; and Weiss, Y., eds.,
  \emph{{{ECCV}} 2018}, volume 11215, 556--572.

\bibitem[{Chen et~al.(2018)Chen, Xu, Zhang, Tang, Cao, Qin, and
  Zha}]{chenSequentialRecommendationUser2018}
Chen, X.; Xu, H.; Zhang, Y.; Tang, J.; Cao, Y.; Qin, Z.; and Zha, H. 2018.
\newblock Sequential {{Recommendation}} with {{User Memory Networks}}.
\newblock In \emph{{{WSDM}} '18}, 108--116.

\bibitem[{Fan et~al.(2021)Fan, Liu, Zhang, Xiong, Zheng, and
  Yu}]{fanContinuousTimeSequentialRecommendation2021}
Fan, Z.; Liu, Z.; Zhang, J.; Xiong, Y.; Zheng, L.; and Yu, P.~S. 2021.
\newblock Continuous-{{Time Sequential Recommendation}} with {{Temporal Graph
  Collaborative Transformer}}.
\newblock In \emph{{{CIKM}} '21}.

\bibitem[{Finn, Abbeel, and
  Levine(2017)}]{finnModelAgnosticMetaLearningFast2017}
Finn, C.; Abbeel, P.; and Levine, S. 2017.
\newblock Model-{{Agnostic Meta}}-{{Learning}} for {{Fast Adaptation}} of
  {{Deep Networks}}.
\newblock In \emph{{{ICML}} '17}, 1126--1135.

\bibitem[{Flennerhag et~al.(2019{\natexlab{a}})Flennerhag, Moreno, Lawrence,
  and Damianou}]{flennerhagTransferringKnowledgeLearning2019}
Flennerhag, S.; Moreno, P.~G.; Lawrence, N.~D.; and Damianou, A.
  2019{\natexlab{a}}.
\newblock Transferring {{Knowledge}} across {{Learning Processes}}.
\newblock In \emph{{{ICLR}}}.

\bibitem[{Flennerhag et~al.(2019{\natexlab{b}})Flennerhag, Rusu, Pascanu,
  Visin, Yin, and Hadsell}]{flennerhagMetaLearningWarpedGradient2019}
Flennerhag, S.; Rusu, A.~A.; Pascanu, R.; Visin, F.; Yin, H.; and Hadsell, R.
  2019{\natexlab{b}}.
\newblock Meta-{{Learning}} with {{Warped Gradient Descent}}.
\newblock In \emph{{{ICLR}}}.

\bibitem[{Goodfellow et~al.(2015)Goodfellow, Mirza, Xiao, Courville, and
  Bengio}]{goodfellowEmpiricalInvestigationCatastrophic2015}
Goodfellow, I.~J.; Mirza, M.; Xiao, D.; Courville, A.; and Bengio, Y. 2015.
\newblock An {{Empirical Investigation}} of {{Catastrophic Forgetting}} in
  {{Gradient}}-{{Based Neural Networks}}.
\newblock \emph{arXiv:1312.6211 [cs, stat]}.

\bibitem[{He et~al.(2020)He, Deng, Wang, Li, Zhang, and
  Wang}]{heLightGCNSimplifyingPowering2020}
He, X.; Deng, K.; Wang, X.; Li, Y.; Zhang, Y.; and Wang, M. 2020.
\newblock {{LightGCN}}: {{Simplifying}} and {{Powering Graph Convolution
  Network}} for {{Recommendation}}.
\newblock In \emph{{{SIGIR}} '20}, 639--648.

\bibitem[{He et~al.(2017)He, Liao, Zhang, Nie, Hu, and
  Chua}]{heNeuralCollaborativeFiltering2017}
He, X.; Liao, L.; Zhang, H.; Nie, L.; Hu, X.; and Chua, T.-S. 2017.
\newblock Neural {{Collaborative Filtering}}.
\newblock In \emph{{{WWW}} '17}, 173--182.

\bibitem[{He et~al.(2016)He, Zhang, Kan, and
  Chua}]{heFastMatrixFactorization2016}
He, X.; Zhang, H.; Kan, M.-Y.; and Chua, T.-S. 2016.
\newblock Fast {{Matrix Factorization}} for {{Online Recommendation}} with
  {{Implicit Feedback}}.
\newblock In \emph{{{SIGIR}} '16}, 549--558.

\bibitem[{Hidasi et~al.(2016)Hidasi, Karatzoglou, Baltrunas, and
  Tikk}]{hidasiSessionbasedRecommendationsRecurrent2016}
Hidasi, B.; Karatzoglou, A.; Baltrunas, L.; and Tikk, D. 2016.
\newblock Session-Based {{Recommendations}} with {{Recurrent Neural Networks}}.
\newblock In \emph{{{ICLR}}}.

\bibitem[{Ji et~al.(2020)Ji, Wang, Wang, Chen, and
  Cristea}]{jiSequentialRecommenderTimeaware2020}
Ji, W.; Wang, K.; Wang, X.; Chen, T.; and Cristea, A. 2020.
\newblock Sequential {{Recommender}} via {{Time}}-Aware {{Attentive Memory
  Network}}.
\newblock In \emph{{{CIKM}} '20}, 565--574.

\bibitem[{Kang and
  McAuley(2018)}]{kangSelfAttentiveSequentialRecommendation2018}
Kang, W.-C.; and McAuley, J. 2018.
\newblock Self-{{Attentive Sequential Recommendation}}.
\newblock In \emph{{{ICDM}} '18}, 197--206.

\bibitem[{Kingma and Ba(2014)}]{kingma2014adam}
Kingma, D.~P.; and Ba, J. 2014.
\newblock Adam: A method for stochastic optimization.
\newblock \emph{arXiv preprint arXiv:1412.6980}.

\bibitem[{Kipf and Welling(2016)}]{kipfSemiSupervisedClassificationGraph2016}
Kipf, T.~N.; and Welling, M. 2016.
\newblock Semi-{{Supervised Classification}} with {{Graph Convolutional
  Networks}}.
\newblock In \emph{{{ICLR}}}.

\bibitem[{Koren(2009)}]{korenCollaborativeFilteringTemporal2009}
Koren, Y. 2009.
\newblock Collaborative Filtering with Temporal Dynamics.
\newblock In \emph{{{KDD}} '09}, 447--456.

\bibitem[{Koren, Bell, and
  Volinsky(2009)}]{korenMatrixFactorizationTechniques2009}
Koren, Y.; Bell, R.; and Volinsky, C. 2009.
\newblock Matrix {{Factorization Techniques}} for {{Recommender Systems}}.
\newblock \emph{Computer}, 42(8): 30--37.

\bibitem[{Lee et~al.(2019)Lee, Im, Jang, Cho, and
  Chung}]{leeMeLUMetaLearnedUser2019}
Lee, H.; Im, J.; Jang, S.; Cho, H.; and Chung, S. 2019.
\newblock {{MeLU}}: {{Meta}}-{{Learned User Preference Estimator}} for
  {{Cold}}-{{Start Recommendation}}.
\newblock In \emph{{{KDD}} '19}, 1073--1082.

\bibitem[{Lee and Choi(2018)}]{leeGradientBasedMetaLearningLearned2018}
Lee, Y.; and Choi, S. 2018.
\newblock Gradient-{{Based Meta}}-{{Learning}} with {{Learned Layerwise
  Metric}} and {{Subspace}}.
\newblock In \emph{{{ICML}} '18}, 2927--2936.

\bibitem[{Li et~al.(2017)Li, Ren, Chen, Ren, Lian, and
  Ma}]{liNeuralAttentiveSessionbased2017}
Li, J.; Ren, P.; Chen, Z.; Ren, Z.; Lian, T.; and Ma, J. 2017.
\newblock Neural {{Attentive Session}}-Based {{Recommendation}}.
\newblock In \emph{{{CIKM}} '17}, 1419--1428.

\bibitem[{Li, Wang, and McAuley(2020)}]{liTimeIntervalAware2020}
Li, J.; Wang, Y.; and McAuley, J. 2020.
\newblock Time {{Interval Aware Self}}-{{Attention}} for {{Sequential
  Recommendation}}.
\newblock In \emph{{{WSDM}} '20}, 322--330.

\bibitem[{Lian et~al.(2021)Lian, Batal, Liu, Soni, Kang, Wang, and
  Xie}]{lianMultiInterestAwareUserModeling2021a}
Lian, J.; Batal, I.; Liu, Z.; Soni, A.; Kang, E.~Y.; Wang, Y.; and Xie, X.
  2021.
\newblock Multi-{{Interest}}-{{Aware User Modeling}} for {{Large}}-{{Scale
  Sequential Recommendations}}.
\newblock \emph{arXiv:2102.09211 [cs]}.

\bibitem[{Liu et~al.(2018)Liu, Zeng, Mokhosi, and
  Zhang}]{liuSTAMPShortTermAttention2018}
Liu, Q.; Zeng, Y.; Mokhosi, R.; and Zhang, H. 2018.
\newblock {{STAMP}}: {{Short}}-{{Term Attention}}/{{Memory Priority Model}} for
  {{Session}}-Based {{Recommendation}}.
\newblock In \emph{{{KDD}} '18}, 1831--1839.

\bibitem[{Liu et~al.(2020)Liu, Chen, Sun, Xie, Gao, Ding, and
  Shen}]{liuIntentPreferenceDecoupling2020}
Liu, Z.; Chen, H.; Sun, F.; Xie, X.; Gao, J.; Ding, B.; and Shen, Y. 2020.
\newblock Intent {{Preference Decoupling}} for {{User Representation}} on
  {{Online Recommender System}}.
\newblock In \emph{{{IJCAI}}-20}, 2575--2582.

\bibitem[{Luo et~al.(2020)Luo, Chen, Cheng, Dong, He, Feng, and
  Li}]{luoMetaSelectorMetaLearningRecommendation2020}
Luo, M.; Chen, F.; Cheng, P.; Dong, Z.; He, X.; Feng, J.; and Li, Z. 2020.
\newblock {{MetaSelector}}: {{Meta}}-{{Learning}} for {{Recommendation}} with
  {{User}}-{{Level Adaptive Model Selection}}.
\newblock In \emph{{{WWW}} '20}, 2507--2513.

\bibitem[{Ma et~al.(2020)Ma, Ma, Zhang, Sun, Liu, and
  Coates}]{maMemoryAugmentedGraph2020}
Ma, C.; Ma, L.; Zhang, Y.; Sun, J.; Liu, X.; and Coates, M. 2020.
\newblock Memory {{Augmented Graph Neural Networks}} for {{Sequential
  Recommendation}}.
\newblock In \emph{{{AAAI}}-20}, volume~34, 5045--5052.

\bibitem[{Ni, Li, and McAuley(2019)}]{niJustifyingRecommendationsUsing2019}
Ni, J.; Li, J.; and McAuley, J. 2019.
\newblock Justifying {{Recommendations}} Using {{Distantly}}-{{Labeled
  Reviews}} and {{Fine}}-{{Grained Aspects}}.
\newblock In \emph{{{EMNLP}}-{{IJCNLP}} 2019}, 188--197.

\bibitem[{Nichol, Achiam, and Schulman(2018)}]{nichol2018firstorder}
Nichol, A.; Achiam, J.; and Schulman, J. 2018.
\newblock On First-Order Meta-Learning Algorithms.
\newblock arXiv:1803.02999.

\bibitem[{Pascanu, Mikolov, and
  Bengio(2013)}]{pascanuDifficultyTrainingRecurrent2013}
Pascanu, R.; Mikolov, T.; and Bengio, Y. 2013.
\newblock On the Difficulty of Training Recurrent Neural Networks.
\newblock In \emph{{{ICML}} '13}, III--1310--III--1318.

\bibitem[{Ravi and Larochelle(2016)}]{raviOptimizationModelFewShot2016}
Ravi, S.; and Larochelle, H. 2016.
\newblock Optimization as a {{Model}} for {{Few}}-{{Shot Learning}}.
\newblock In \emph{{{ICLR}}}.

\bibitem[{Rendle et~al.(2009)Rendle, Freudenthaler, Gantner, and
  {Schmidt-Thieme}}]{rendleBPRBayesianPersonalized2009}
Rendle, S.; Freudenthaler, C.; Gantner, Z.; and {Schmidt-Thieme}, L. 2009.
\newblock {{BPR}}: {{Bayesian}} Personalized Ranking from Implicit Feedback.
\newblock In \emph{{{UAI}} '09}, 452--461.

\bibitem[{Ricci, Rokach, and
  Shapira(2011)}]{ricciIntroductionRecommenderSystems2011}
Ricci, F.; Rokach, L.; and Shapira, B. 2011.
\newblock Introduction to {{Recommender Systems Handbook}}.
\newblock In Ricci, F.; Rokach, L.; Shapira, B.; and Kantor, P.~B., eds.,
  \emph{Recommender {{Systems Handbook}}}, 1--35. {Springer US}.

\bibitem[{Rossi et~al.(2020)Rossi, Chamberlain, Frasca, Eynard, Monti, and
  Bronstein}]{tgn_icml_grl2020}
Rossi, E.; Chamberlain, B.; Frasca, F.; Eynard, D.; Monti, F.; and Bronstein,
  M. 2020.
\newblock Temporal Graph Networks for Deep Learning on Dynamic Graphs.
\newblock In \emph{{{ICML}} 2020 Workshop on Graph Representation Learning}.

\bibitem[{Santoro et~al.(2016)Santoro, Bartunov, Botvinick, Wierstra, and
  Lillicrap}]{santoroMetaLearningMemoryAugmentedNeural2016}
Santoro, A.; Bartunov, S.; Botvinick, M.; Wierstra, D.; and Lillicrap, T. 2016.
\newblock Meta-{{Learning}} with {{Memory}}-{{Augmented Neural Networks}}.
\newblock In \emph{{{ICML}} '16}, 1842--1850.

\bibitem[{Scarselli et~al.(2009)Scarselli, Gori, Tsoi, Hagenbuchner, and
  Monfardini}]{scarselliGraphNeuralNetwork2009}
Scarselli, F.; Gori, M.; Tsoi, A.~C.; Hagenbuchner, M.; and Monfardini, G.
  2009.
\newblock The {{Graph Neural Network Model}}.
\newblock \emph{IEEE Transactions on Neural Networks}, 20(1): 61--80.

\bibitem[{Schafer, Konstan, and
  Riedl(1999)}]{schaferRecommenderSystemsEcommerce1999}
Schafer, J.~B.; Konstan, J.; and Riedl, J. 1999.
\newblock Recommender Systems in E-Commerce.
\newblock In \emph{{{EC}} '99}, 158--166.

\bibitem[{Sun et~al.(2019)Sun, Liu, Wu, Pei, Lin, Ou, and
  Jiang}]{sunBERT4RecSequentialRecommendation2019}
Sun, F.; Liu, J.; Wu, J.; Pei, C.; Lin, X.; Ou, W.; and Jiang, P. 2019.
\newblock {{BERT4Rec}}: {{Sequential Recommendation}} with {{Bidirectional
  Encoder Representations}} from {{Transformer}}.
\newblock In \emph{{{CIKM}} '19}, 1441--1450.

\bibitem[{Truong, Salah, and
  Lauw(2021)}]{truongBilateralVariationalAutoencoder2021}
Truong, Q.-T.; Salah, A.; and Lauw, H.~W. 2021.
\newblock Bilateral {{Variational Autoencoder}} for {{Collaborative
  Filtering}}.
\newblock In \emph{{{WSDM}} '21}, 292--300.

\bibitem[{Vartak et~al.(2017)Vartak, Thiagarajan, Miranda, Bratman, and
  Larochelle}]{vartakMetalearningPerspectiveColdstart2017}
Vartak, M.; Thiagarajan, A.; Miranda, C.; Bratman, J.; and Larochelle, H. 2017.
\newblock A Meta-Learning Perspective on Cold-Start Recommendations for Items.
\newblock In \emph{{{NIPS}}'17}, 6907--6917.

\bibitem[{Vaswani et~al.(2017)Vaswani, Shazeer, Parmar, Uszkoreit, Jones,
  Gomez, Kaiser, and Polosukhin}]{vaswaniAttentionAllYou2017}
Vaswani, A.; Shazeer, N.; Parmar, N.; Uszkoreit, J.; Jones, L.; Gomez, A.~N.;
  Kaiser, {\L}.; and Polosukhin, I. 2017.
\newblock Attention Is {{All}} You {{Need}}.
\newblock In \emph{{{NIPS}} '17}, volume~30.

\bibitem[{Veli{\v c}kovi{\'c} et~al.(2018)Veli{\v c}kovi{\'c}, Cucurull,
  Casanova, Romero, Li{\`o}, and Bengio}]{velickovicGraphAttentionNetworks2018}
Veli{\v c}kovi{\'c}, P.; Cucurull, G.; Casanova, A.; Romero, A.; Li{\`o}, P.;
  and Bengio, Y. 2018.
\newblock Graph {{Attention Networks}}.
\newblock In \emph{{{ICLR}}}.

\bibitem[{Vilalta and Drissi(2002)}]{vilaltaPerspectiveViewSurvey2002}
Vilalta, R.; and Drissi, Y. 2002.
\newblock A {{Perspective View}} and {{Survey}} of {{Meta}}-{{Learning}}.
\newblock \emph{Artificial Intelligence Review}, 18(2): 77--95.

\bibitem[{Wang and Caverlee(2019)}]{wangRecurrentRecommendationLocal2019}
Wang, J.; and Caverlee, J. 2019.
\newblock Recurrent {{Recommendation}} with {{Local Coherence}}.
\newblock In \emph{{{WSDM}} '19}, 564--572.

\bibitem[{Wang, Ding, and
  Caverlee(2021)}]{wangSequentialRecommendationColdstart2021}
Wang, J.; Ding, K.; and Caverlee, J. 2021.
\newblock Sequential {{Recommendation}} for {{Cold}}-Start {{Users}} with
  {{Meta Transitional Learning}}.
\newblock In \emph{{{SIGIR}} '21}, 1783--1787.

\bibitem[{Wang et~al.(2020)Wang, Ding, Hong, Liu, and
  Caverlee}]{wangNextitemRecommendationSequential2020}
Wang, J.; Ding, K.; Hong, L.; Liu, H.; and Caverlee, J. 2020.
\newblock Next-Item {{Recommendation}} with {{Sequential Hypergraphs}}.
\newblock In \emph{{{SIGIR}} '20}, 1101--1110.

\bibitem[{Wang et~al.(2018{\natexlab{a}})Wang, Yin, Hu, Lian, Wang, and
  Huang}]{wangNeuralMemoryStreaming2018}
Wang, Q.; Yin, H.; Hu, Z.; Lian, D.; Wang, H.; and Huang, Z.
  2018{\natexlab{a}}.
\newblock Neural {{Memory Streaming Recommender Networks}} with {{Adversarial
  Training}}.
\newblock In \emph{{{KDD}} '18}, 2467--2475.

\bibitem[{Wang et~al.(2019{\natexlab{a}})Wang, Hu, Wang, Cao, Sheng, and
  Orgun}]{wangSequentialRecommenderSystems2019a}
Wang, S.; Hu, L.; Wang, Y.; Cao, L.; Sheng, Q.~Z.; and Orgun, M.
  2019{\natexlab{a}}.
\newblock Sequential {{Recommender Systems}}: {{Challenges}}, {{Progress}} and
  {{Prospects}}.
\newblock In \emph{{{IJCAI}}-19}, 6332--6338.

\bibitem[{Wang et~al.(2018{\natexlab{b}})Wang, Yin, Huang, Wang, Du, and
  Nguyen}]{wangStreamingRankingBased2018}
Wang, W.; Yin, H.; Huang, Z.; Wang, Q.; Du, X.; and Nguyen, Q. V.~H.
  2018{\natexlab{b}}.
\newblock Streaming {{Ranking Based Recommender Systems}}.
\newblock In \emph{{{SIGIR}} '18}, 525--534.

\bibitem[{Wang et~al.(2019{\natexlab{b}})Wang, He, Wang, Feng, and
  Chua}]{wangNeuralGraphCollaborative2019}
Wang, X.; He, X.; Wang, M.; Feng, F.; and Chua, T.-S. 2019{\natexlab{b}}.
\newblock Neural {{Graph Collaborative Filtering}}.
\newblock In \emph{{{SIGIR}} '19}, 165--174.

\bibitem[{Wei et~al.(2020)Wei, Wu, Li, Hu, Feng, He, Sun, and
  Wang}]{weiFastAdaptationColdStart2020}
Wei, T.; Wu, Z.; Li, R.; Hu, Z.; Feng, F.; He, X.; Sun, Y.; and Wang, W. 2020.
\newblock Fast {{Adaptation}} for {{Cold}}-{{Start Collaborative Filtering}}
  with {{Meta}}-{{Learning}}.
\newblock In \emph{{{ICDM}} '20}, 661--670.

\bibitem[{Weston, Chopra, and Bordes(2015)}]{westonMemoryNetworks2015}
Weston, J.; Chopra, S.; and Bordes, A. 2015.
\newblock Memory {{Networks}}.
\newblock In \emph{{{ICLR}}}.

\bibitem[{Wu et~al.(2017)Wu, Ahmed, Beutel, Smola, and
  Jing}]{wuRecurrentRecommenderNetworks2017}
Wu, C.-Y.; Ahmed, A.; Beutel, A.; Smola, A.~J.; and Jing, H. 2017.
\newblock Recurrent {{Recommender Networks}}.
\newblock In \emph{{{WSDM}} '17}, 495--503.

\bibitem[{Xie et~al.(2021)Xie, Wang, Wang, Lu, Zou, Xia, and
  Lin}]{xieLongShortTermTemporal2021}
Xie, R.; Wang, Y.; Wang, R.; Lu, Y.; Zou, Y.; Xia, F.; and Lin, L. 2021.
\newblock Long {{Short}}-{{Term Temporal Meta}}-Learning in {{Online
  Recommendation}}.
\newblock \emph{arXiv:2105.03686 [cs]}.

\bibitem[{Xiong et~al.(2010)Xiong, Chen, Huang, Schneider, and
  Carbonell}]{xiongTemporalCollaborativeFiltering2010}
Xiong, L.; Chen, X.; Huang, T.-K.; Schneider, J.; and Carbonell, J.~G. 2010.
\newblock Temporal {{Collaborative Filtering}} with {{Bayesian Probabilistic
  Tensor Factorization}}.
\newblock In \emph{{{SIAM}} '10}, 211--222.

\bibitem[{Xu et~al.(2019)Xu, Ruan, Korpeoglu, Kumar, and
  Achan}]{xuInductiveRepresentationLearning2019}
Xu, D.; Ruan, C.; Korpeoglu, E.; Kumar, S.; and Achan, K. 2019.
\newblock Inductive Representation Learning on Temporal Graphs.
\newblock In \emph{{{ICLR}}}.

\bibitem[{Ye et~al.(2020)Ye, Wang, Chen, Wang, Qin, and
  Yin}]{yeTimeMattersSequential2020}
Ye, W.; Wang, S.; Chen, X.; Wang, X.; Qin, Z.; and Yin, D. 2020.
\newblock Time {{Matters}}: {{Sequential Recommendation}} with {{Complex
  Temporal Information}}.
\newblock In \emph{{{SIGIR}} '20}, 1459--1468.

\bibitem[{Yu et~al.(2019{\natexlab{a}})Yu, Si, Hu, and
  Zhang}]{yuReviewRecurrentNeural2019}
Yu, Y.; Si, X.; Hu, C.; and Zhang, J. 2019{\natexlab{a}}.
\newblock A {{Review}} of {{Recurrent Neural Networks}}: {{LSTM Cells}} and
  {{Network Architectures}}.
\newblock \emph{Neural Computation}, 31(7): 1235--1270.

\bibitem[{Yu et~al.(2019{\natexlab{b}})Yu, Lian, Mahmoody, Liu, and
  Xie}]{yuAdaptiveUserModeling2019}
Yu, Z.; Lian, J.; Mahmoody, A.; Liu, G.; and Xie, X. 2019{\natexlab{b}}.
\newblock Adaptive User Modeling with Long and Short-Term Preferences for
  Personalized Recommendation.
\newblock In \emph{{{IJCAI}}-19}, 4213--4219.

\bibitem[{Zhang et~al.(2020)Zhang, Feng, Wang, He, Wang, Li, and
  Zhang}]{zhangHowRetrainRecommender2020}
Zhang, Y.; Feng, F.; Wang, C.; He, X.; Wang, M.; Li, Y.; and Zhang, Y. 2020.
\newblock How to {{Retrain Recommender System}}? {{A Sequential
  Meta}}-{{Learning Method}}.
\newblock In \emph{{{SIGIR}} '20}, 1479--1488.

\bibitem[{Zhu et~al.(2017)Zhu, Li, Liao, Wang, Guan, Liu, and
  Cai}]{zhuWhatNextModeling2017}
Zhu, Y.; Li, H.; Liao, Y.; Wang, B.; Guan, Z.; Liu, H.; and Cai, D. 2017.
\newblock What to {{Do Next}}: {{Modeling User Behaviors}} by
  {{Time}}-{{LSTM}}.
\newblock In \emph{{{IJCAI}}-17}, 3602--3608.

\end{thebibliography}

\newpage

\appendix

\section{Implementation Details}

We implemented LeapRec based on PyTorch\footnote{\url{https://pytorch.org/}} and PyTorch Geometric\footnote{\url{https://pytorch-geometric.readthedocs.io/}} libraries in Python 3.6.
All experiments except for TGN and MetaTL were conducted on Intel Xeon Gold 6148 CPU @ 2.4 GHz, 5 Cores, 60GB RAM, NVIDIA Tesla V100-16GB.
The experiments for TGN and MetaTL were conducted on Intel Xeon Platinum 8360Y CPU @ 2.4 GHz, 9 Cores, 60GB RAM, NVIDIA A100-40GB.

\section{Additional Ablation Study}

We investigate the effect of LeapRec on the neural recommender by introducing ``without LeapRec'' variants: the neural recommender alone (--LeapRec) and with positional encoding for the self-attention (--LeapRec+Pos.SA).
We conducted additional experiments for those variants by using the same setting in Section~\ref{sec:experiments} of the main paper.

\begin{table*}[ht]
    \centering
    \small
      \begin{adjustbox}{max width=\textwidth}
      \begin{tabular}{lcccc cccccccc}
      \toprule
      \multirow{2}[4]{*}{Neural Recommender} & \multicolumn{4}{c}{Amazon} & \multicolumn{4}{c}{Goodreads} & \multicolumn{4}{c}{Yelp} \\
    
      \cmidrule(lr){2-5} \cmidrule(lr){6-9} \cmidrule(lr){10-13}      & \scriptsize{HR@1}  & \scriptsize{HR@5}  & \scriptsize{NDCG@5}  & \scriptsize{MRR} & \scriptsize{HR@1}  & \scriptsize{HR@5}  & \scriptsize{NDCG@5}  & \scriptsize{MRR} & \scriptsize{HR@1}  & \scriptsize{HR@5}  & \scriptsize{NDCG@5}  & \scriptsize{MRR} \\
      \midrule
    
    --LeapRec & 0.0871 & 0.2497 & 0.1696 & 0.1815 & 0.2256 & 0.5470 & 0.3920 & 0.3764 & 0.3151 & 0.6997 & 0.5166 & 0.4841 \\
    --LeapRec+Pos.SA & 0.0962 & 0.2759 & 0.1873 & 0.1980 & 0.2180 & 0.5376 & 0.3836 & 0.3688 & 0.3191 & 0.7040 & 0.5212 & 0.4887 \\
    \textbf{+LeapRec} (Default) & \textbf{0.1252} & \textbf{0.3484} & \textbf{0.2391} & \textbf{0.2443} & \textbf{0.2452} & \textbf{0.5760} & \textbf{0.4171} & \textbf{0.3991} & \textbf{0.3420} & \textbf{0.7466} & \textbf{0.5547} & \textbf{0.5165} \\
    \cmidrule(lr){2-5} \cmidrule(lr){6-9} \cmidrule(lr){10-13} 
    \emph{Improvement} & 30.13\% & 26.30\% & 27.69\% & 23.39\% & 8.65\% & 5.31\% & 6.40\% & 6.03\% & 7.18\% & 6.06\% & 6.42\% & 5.67\% \\
      \bottomrule
      \end{tabular}%
    \end{adjustbox}
      
    \caption{Ablation study of LeapRec. \emph{Improvement} indicates \emph{relative} improvement of \textbf{+LeapRec} over the other best performing variant in each case.}      
    \label{tab:exp-abla}%
  \end{table*}%

We report the results in Table~\ref{tab:exp-abla}. We observe that the neural recommender with LeapRec (+LeapRec) significantly outperforms other variants without LeapRec by a large margin. The results demonstrate the important of LeapRec to the nueral recommender, which cannot simply achieve with positional encoding.

\section{Theoretical Analyses of LeapRec}

In the main part of our paper, we have presented a strong empirical evidence for the effectiveness of LeapRec in the recommendation context.
Due to the space limit of the main paper, we discuss the theoretical implications of our designs in this section of the Appendix.
We first recall some theoretical background and assumptions of our work.

\subsection{Theoretical Preliminaries}

Let $f(\theta): \Theta \rightarrow \mathbb{R}^d$ be a differentiable learning objective parameterized by a set of parameters $\theta \in \Theta$ (often these are learning model parameters). 
Meta-learning theory is often developed under the manifold learning theory framework. 
In this geometric view, a point $p$ is defined as $p := (\theta, f(\theta))$ and the vector space local to each point is assumed to be homeomorphic to a Euclidean space.
Given a task, or in our work a recommendation problem in time period $t$, we can define a task manifold $M$ such that $(\theta, f(\theta)) \in M$.
The local vector space to $p$ is denoted by $T_p M$, it is also known as the \emph{tangent space}.
Using these definitions, we can formally define a \emph{gradient path} of task $t$ as a curve function $\rho: [0,1] \rightarrow M_t$ where $\rho(k) = (\theta^k, f(\theta^k)$), where $i$ denotes the step parameter for gradient flow.
Although the value of $k$ in this definition is in $[0,1]$, it is trivial to see that $k$ can be scaled to represent integer gradient steps in practice.
The remaining of this section assumes a maximum number of gradient steps $K$ and the values of $k$ ranges from 0 to $K$.

\begin{figure}[H]
    \centering
    \includegraphics[width=.6\linewidth]{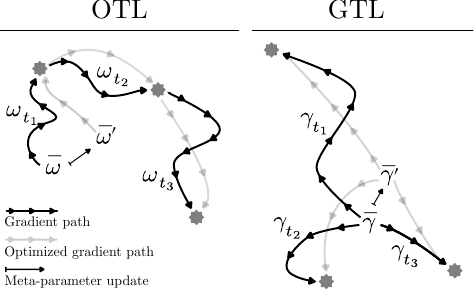}
    \caption{Illustration of OTL and GTL algorithms. The GTL (right) is often used in few-shot learning literature.}
    \label{fig:otl_gtl}
\end{figure}

Given the description of gradient process, the cummulative chordal distance~\citep{ahlberg1967theory} is given as follows.
\begin{align}
    \label{eq:dlength}
    d_2(\theta^0, M) = \sum_{k=0}^{K-1} ||\rho(k+1) - \rho(k)||_2^2
\end{align}
Let $\mu$ be an updating rule such that $\mu(\theta^{k}) = \theta^{k+1}$. 
Eq.~\ref{eq:dlength} is the discretized distance from $\theta^0$ to $\theta^K$, which approximates the length of the true gradient path.
In Figure~\ref{fig:otl_gtl} (also Figure~\ref{fig:overview} in the main paper), the gradient paths are represented by lines with arrows and meta updates minimize the lengths of these gradient paths.

\subsection{Global Time Leap}

The following proposition is based on Theorem 1 of~\citet{flennerhagTransferringKnowledgeLearning2019}. 

\begin{proposition}[Global Time Leap]
\label{prop:equivalent-optimization}
Assuming the data from each time step is drawn i.i.d. from a distribution: $\mathcal{D}_t \sim p_{\mathcal{D}}$,
and there exist a set of parameters such that if $\theta^0$ belongs to such a set,
then GTL yields the solution $\GM$ to the following optimization problem:
\begin{align}
   \min_{\GM} \ &\mathbb{E}_{\mathcal{D}_t \sim p(\mathcal{D})}[d_2(\GM, M_t)], \\ \nonumber
   \text{s.t. } & \GM^{i+1}_t = \mu_t(\GM^i), \GM^0_t = \GM^0, \\
   & \GM \in \Theta = \cap_t \{\GM \ \vert \ f_t(\theta_t^K) \leq f_t(\gamma^{K-1}) \} \nonumber
\end{align}
where the manifold $M_t$ induced by each task $t$.
\end{proposition}

This proposition shows that the gradient updating mechanism of GTL is equivalent with solving the minimum gradient path problem. 
The optimal initialization point $\GM$ is the Pareto optimal initialization with respect to the time steps.
Intuitively, the obtained $\GM$ contains information from each time step by considering them equally.
Hence, GTL captures the global time information.
As we show in Section~\ref{sec:experiments}, GTL is the main predictor in datasets with less varying preferences such as online book repositories.
\\

\newcommand*{\QED}{\hfill\ensuremath{\blacksquare}}
\noindent
\emph{Proof of Proposition~\ref{prop:equivalent-optimization}}.
Let $F_{\GM} = \mathbb{E}_{t \sim p(t)}[d_2(\GM,M_t)]$, the gradient of $F_{\GM}$ with respect to $\GM$ is given by
\begin{align}
    \nabla F_{\GM} = 2 \mathbb{E}_{t \sim p(t)} \left[ \sum_{k=0}^{K-1} - J_t^k(\GM_t) ( \Delta f^k_t \nabla f_t (\gamma_t^k) + \Delta \gamma^k_{\tau}) \right] \nonumber
\end{align}
% TODO: Fix sign of this in the final version

This equation is equivalent to Eq.~\ref{eq:leap:meta_grad} in the main part of the paper.
By approximating the Jacobian matrix $J^k_t \approx I$, we obtain Eq.~\ref{eq:leap:meta_grad}.
In Algorithm~\ref{alg:training}, each iteration (line 4 to 11) accumulates the gradient paths $\{\gamma_t^k\}_{k=0\dots K}$ and line 13 update $\gamma$ to minimize the expected length $F_{\GM}$. \QED

Furthermore, by~\citep[Theorem 1]{flennerhagTransferringKnowledgeLearning2019}, assuming $\Theta$ is a non-empty set and $\beta$ is sufficiently small (line 13 of Algorithm~\ref{alg:training}), there exists a learning rate schedules $\{\alpha^k_t\}_{k=1}^K$ such that Algorithm~\ref{alg:training} converges at a limit point within $\Theta$.
This fact is supported by our empirical analysis of the loss convergence across time steps in (RQ4).

\subsection{Ordered Time Leap}

The main different between GTL and OTL is that OLT has a different gradient updating rules (see Section~\ref{ssec:leap-time}).
Hence, the minimization process optimizes a total gradient path going through each optimal points (see the difference between $\OM$ and $\GM$ optimization in Figure~\ref{fig:otl_gtl}).
Using the same proof technique for GTL, we can show that Algorithm~\ref{alg:training} optimizes the length of the following gradient paths.

\begin{proposition}[Ordered Time Leap]
\label{prop:otl}
Having the same assumptions as Proposition~\ref{prop:equivalent-optimization},
OTL yields the solution to the following optimization problem:
\begin{align}
   \min_{\OM} \sum_{t=1}^T d_2^t(\Lparam{\OT}{t}{K}, M_t),
\end{align}
where $d_2^t(\Lparam{\OT}{t}{K}, M_\mathcal{L})$ is the distance between $\Lparam{\OT}{t}{K}$ and $\Lparam{\OT}{t-1}{K}$ on the manifold $M_t$.
\end{proposition}

The intuition behind OTL that this technique captures the local time information.
Instead of minimizing to be the ``middle point'' like GTL, OTL minimizes each segment of the gradient path.
In do so, OTL is biased toward more recent data more than the past; hence the short-term learning effect.
In combination, OTL and the aforementioned GTL are natural candidates to replace the short-term and long-term time designs of existing methods.

\section{Additional Figures}
\begin{figure}[H]
    \centering
    \begin{subfigure}{\linewidth}
    \centering
        \includegraphics[width=.7\linewidth]{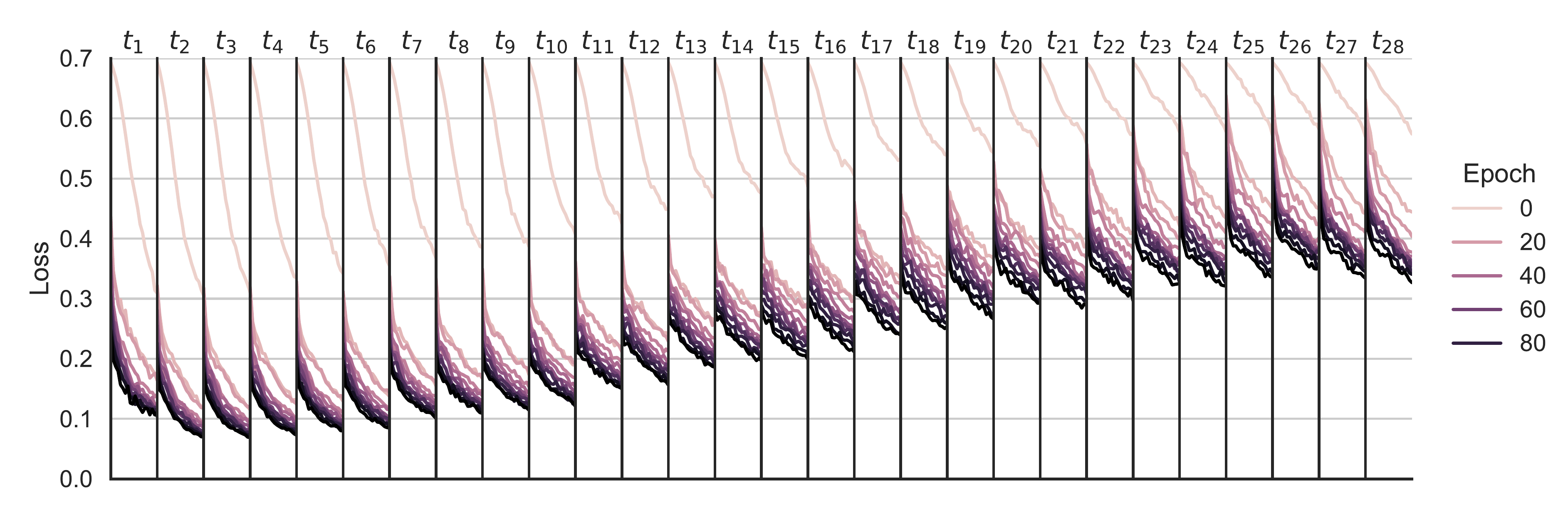}
        \caption{GTL}
    \end{subfigure}
    \begin{subfigure}{\linewidth}
    \centering
        \includegraphics[width=.7\linewidth]{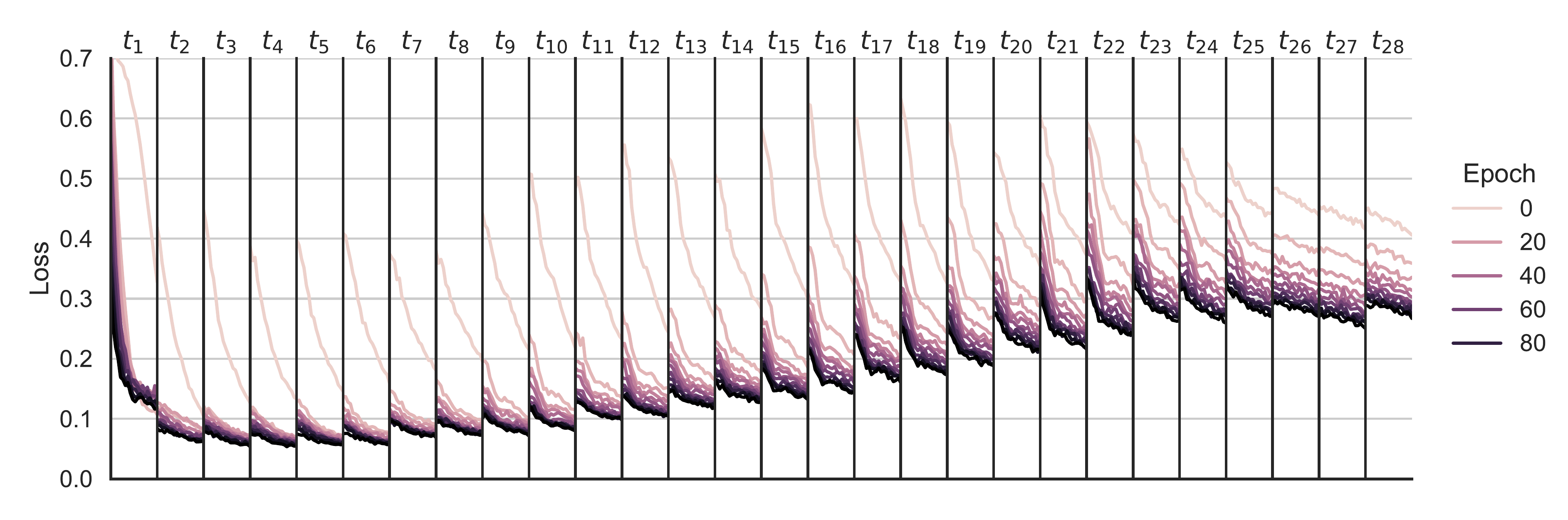}
        \caption{OTL}
    \end{subfigure}    
    \caption{Training loss of individual OTL and GTL in each timestamp on Amazon dataset in different epochs. The horizontal axis in each time window is the gradient-based update steps.\\[-2.0em]}
    \label{fig:training:individual}
\end{figure}

\begin{figure}[H]
    \centering
    \begin{subfigure}{.3\linewidth}
        \includegraphics[width=\linewidth]{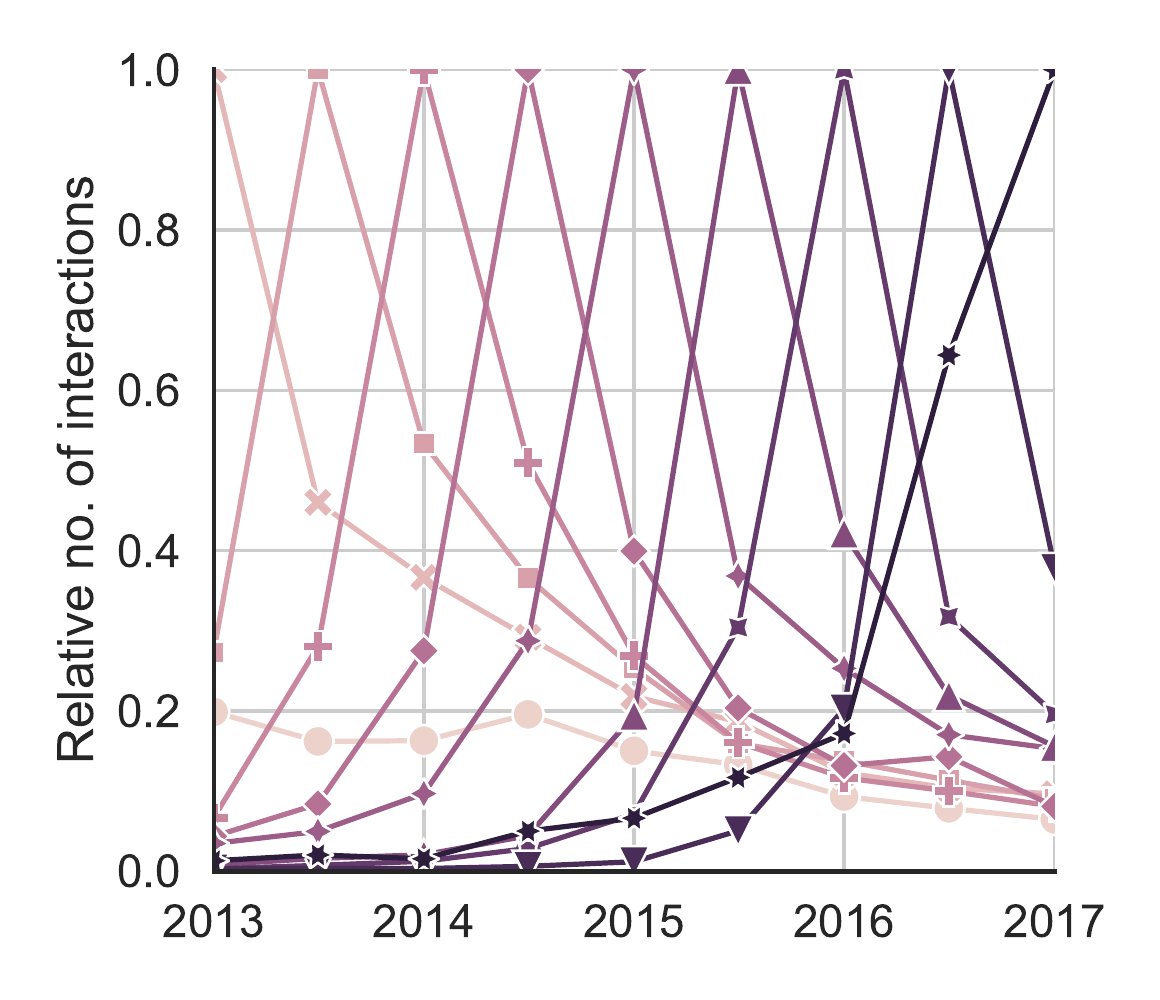}
        \caption{Amazon}
        \label{fig:dataset:top-interactions:amz:apd}
    \end{subfigure}
    \begin{subfigure}{.3\linewidth}
        \includegraphics[width=\linewidth]{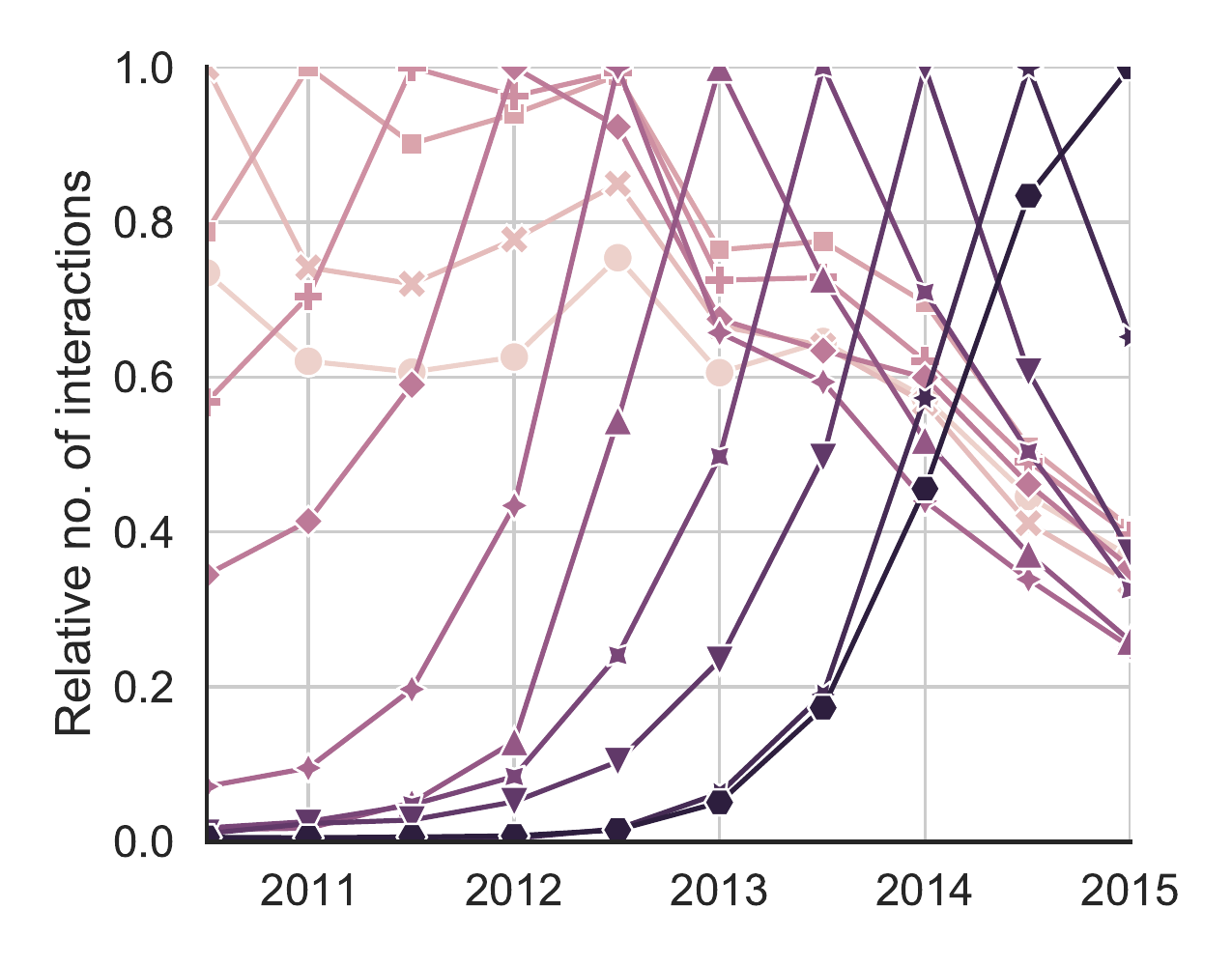}
        \caption{Goodreads}
        \label{fig:dataset:top-interactions:goodreads:apd}
    \end{subfigure}
    \hfill
    \begin{subfigure}{.6\linewidth}
        \includegraphics[width=\linewidth]{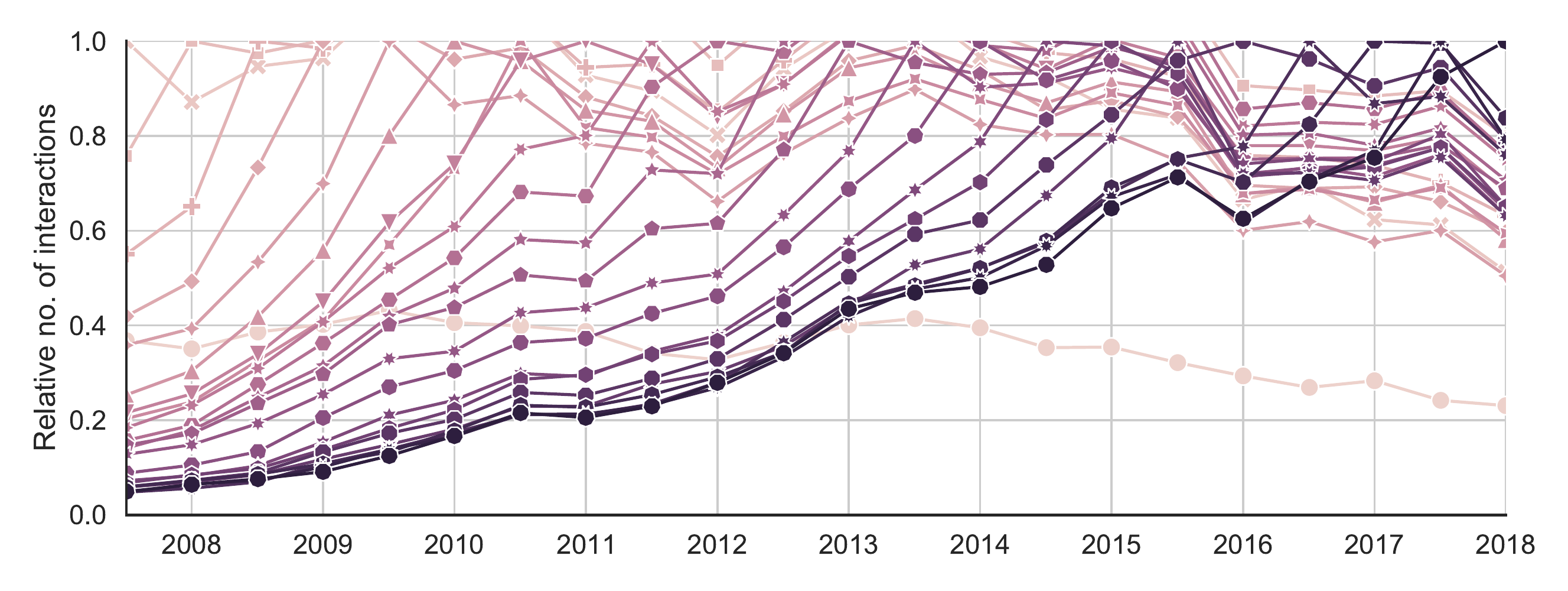}
        \caption{Yelp}
        \label{fig:dataset:top-interactions:yelp:apd}
    \end{subfigure}
    \caption{Relative number of interactions of groups of popular items (top 100) under each timestamp. Each line is a group of items that became popular at the same time.}
    \label{fig:dataset:top-interactions:apd}
\end{figure}

\begin{figure}[H]
    \centering
    \begin{subfigure}{.3\linewidth}
        \includegraphics[width=\linewidth]{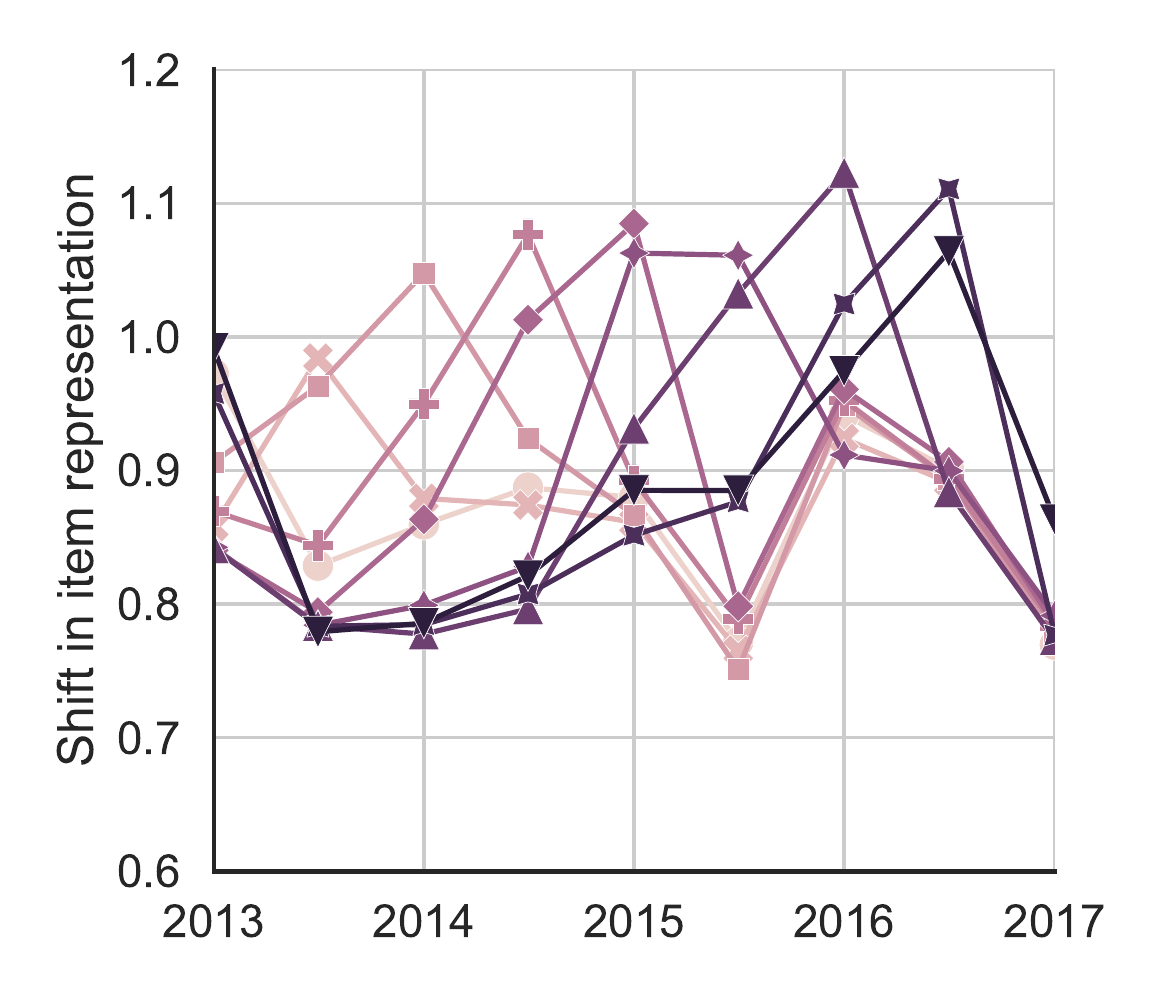}
        \caption{OTL - Amazon}
    \end{subfigure}
    \begin{subfigure}{.3\linewidth}
        \includegraphics[width=\linewidth]{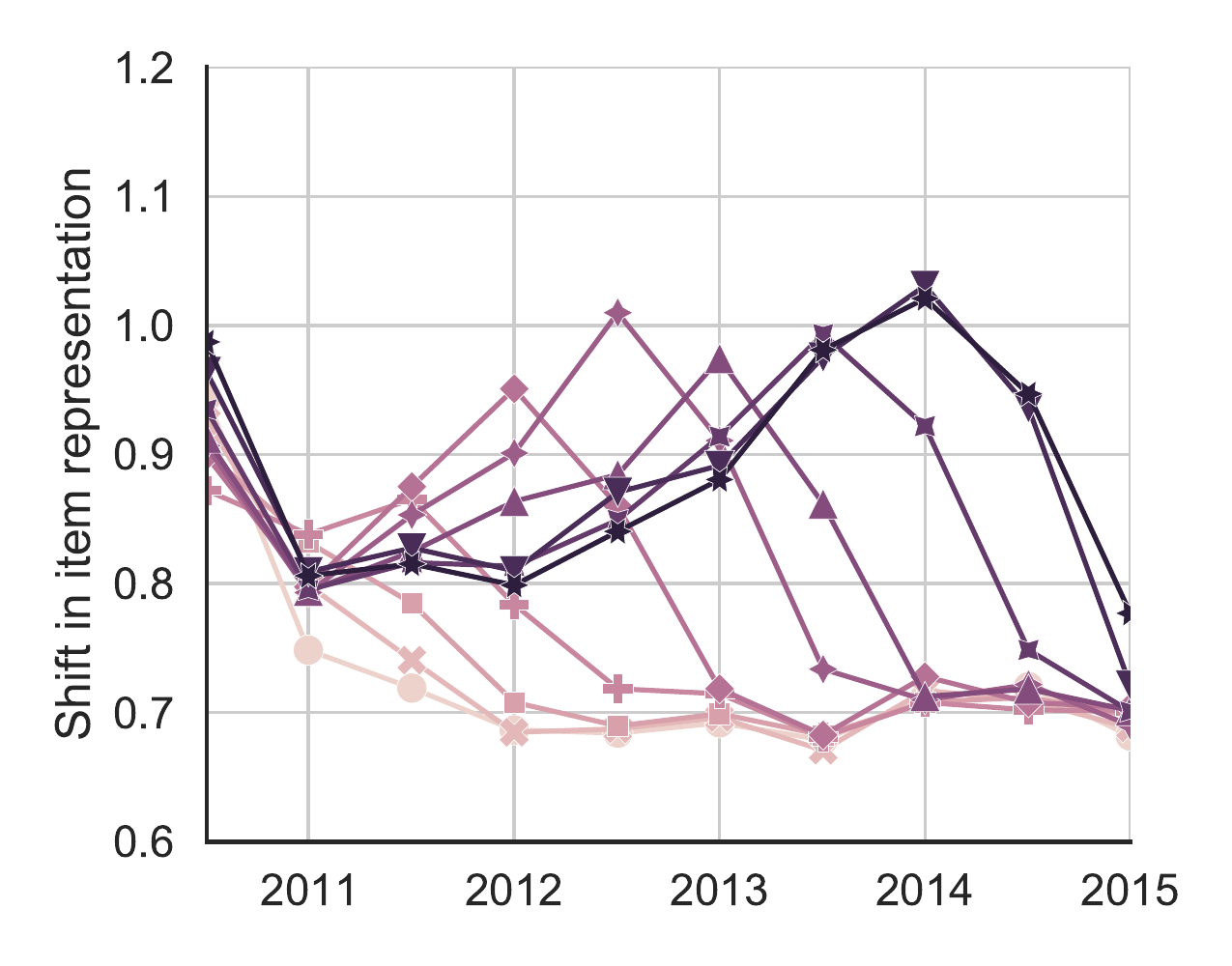}
        \caption{OTL - Goodreads}
    \end{subfigure}
    \hfill
    \centering
    \begin{subfigure}{.6\linewidth}
        \includegraphics[width=\linewidth]{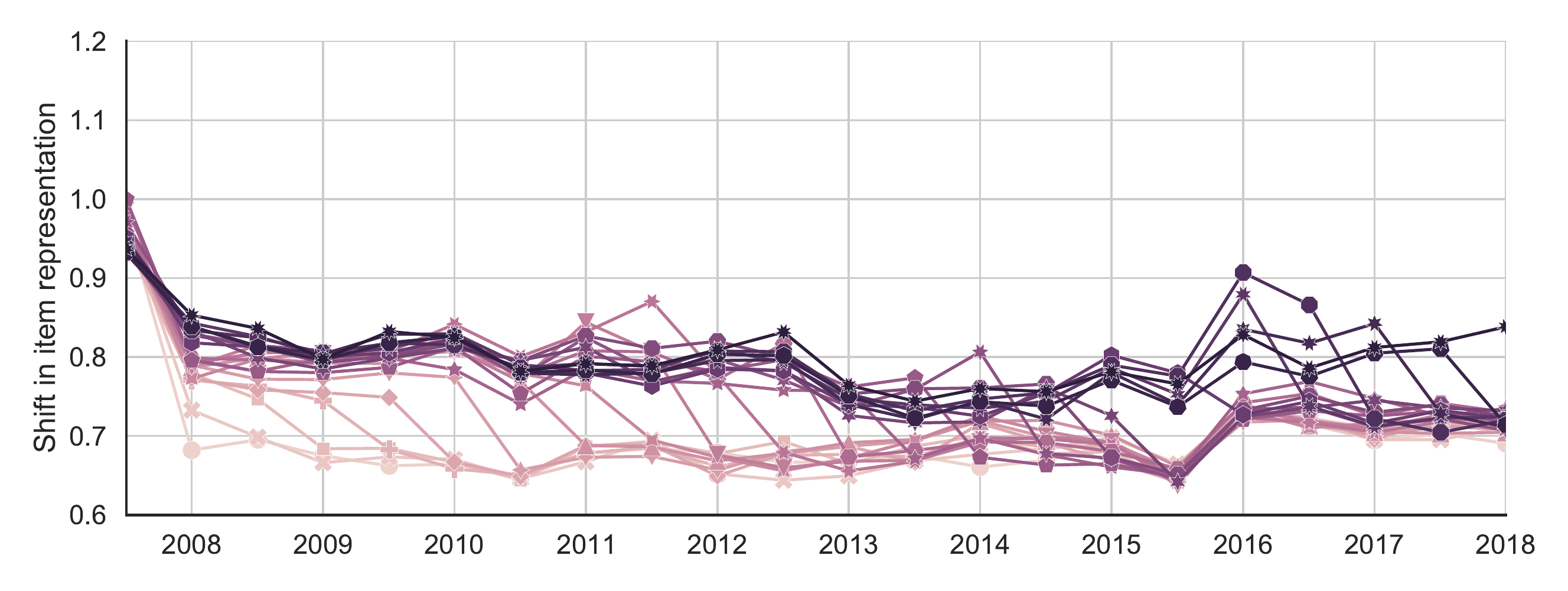}
        \caption{OTL - Yelp}
    \end{subfigure}
    \hfill
    \begin{subfigure}{.4\linewidth}
        \includegraphics[width=.75\linewidth,right]{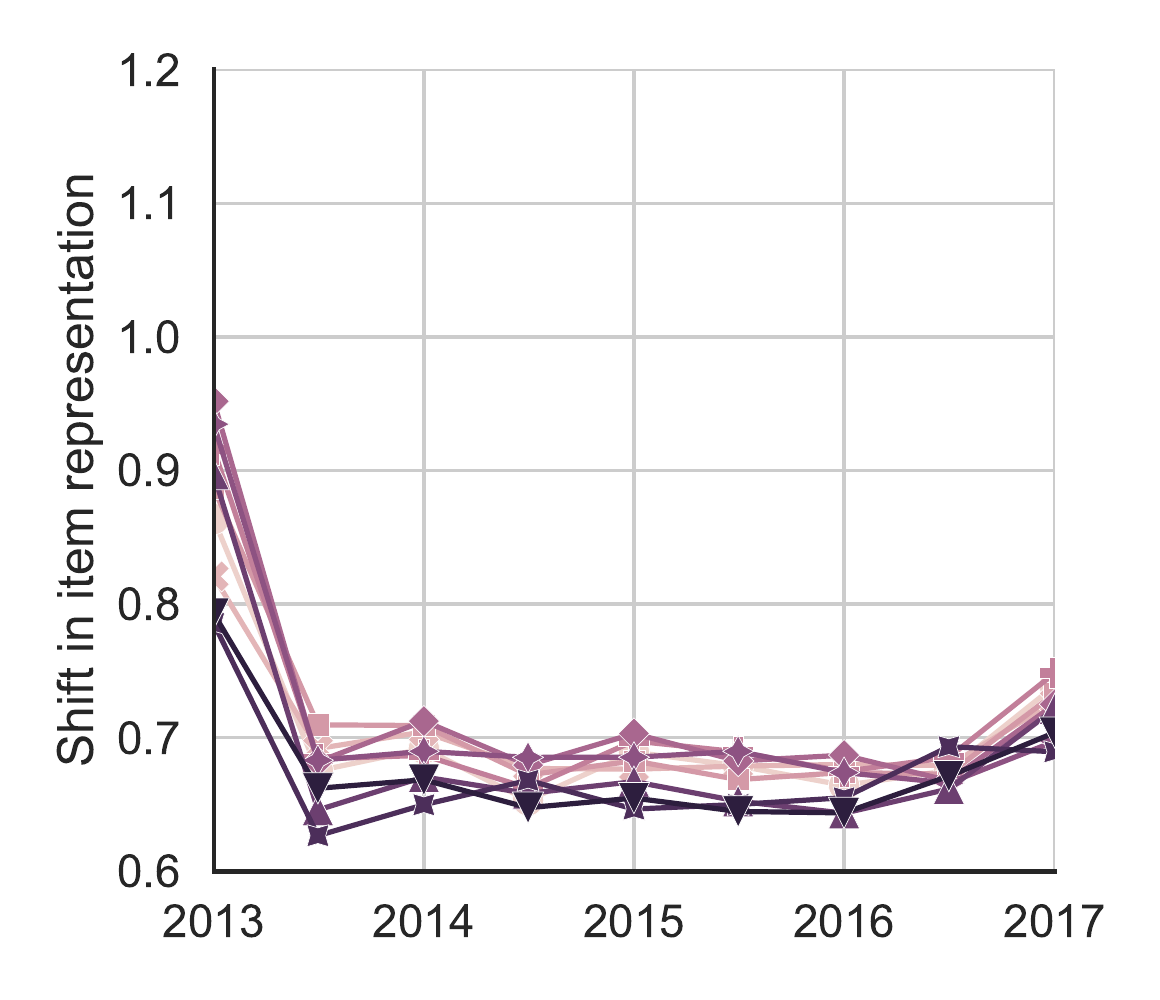}
        \caption{GTL - Amazon}
    \end{subfigure}
    \begin{subfigure}{.4\linewidth}
        \includegraphics[width=.75\linewidth,left]{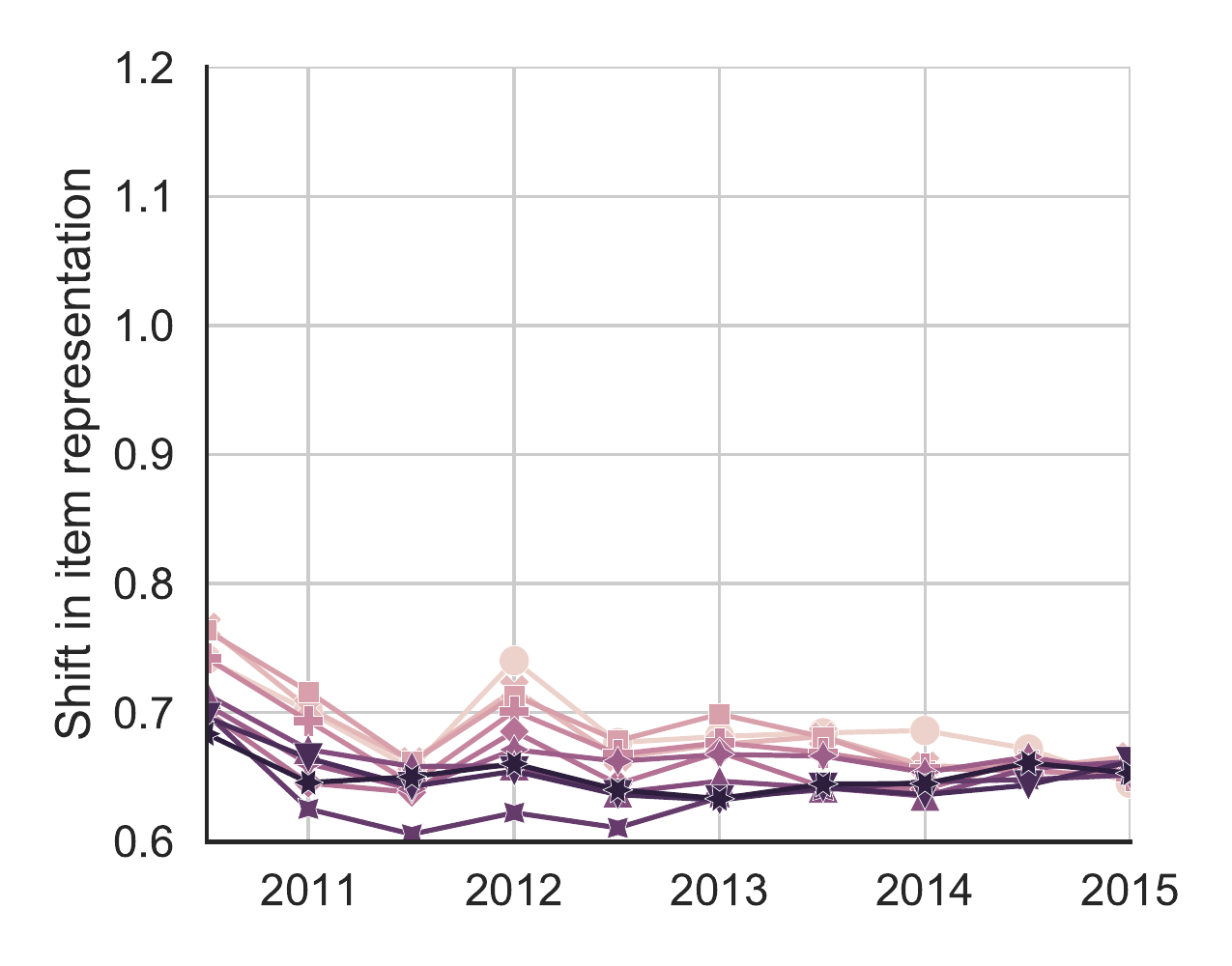}
        \caption{GTL - Goodreads}
    \end{subfigure}
    \hfill
    \begin{subfigure}{.6\linewidth}
        \includegraphics[width=\linewidth]{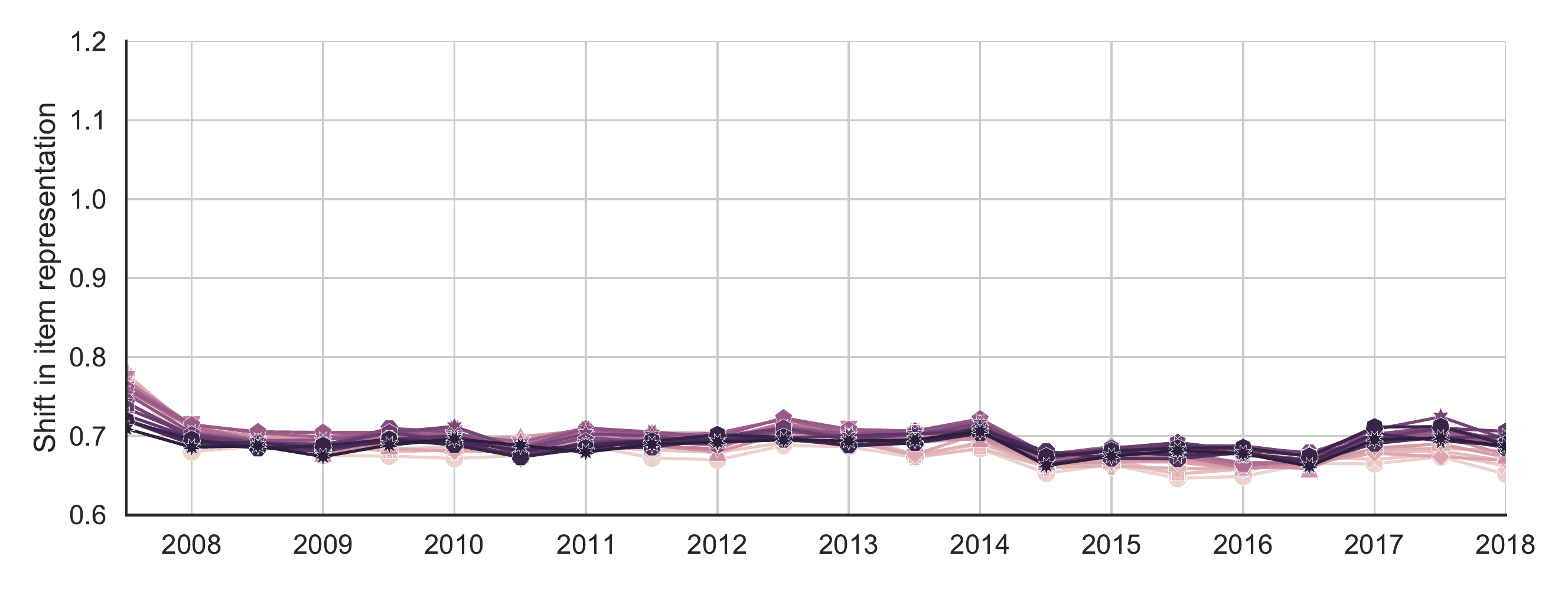}
        \caption{GTL - Yelp}
    \end{subfigure}
    \hfill
    \caption{Shift in the item representations monitored on OTL and GTL of popular items under each timestamp. Each line represents a group of items that became popular at the same time (corresponding to the same group in Figure~\ref{fig:dataset:top-interactions:apd})}
    \label{fig:embedding:apd}
\end{figure}

In Addition, we investigate the convergence of individual GTL and OTL by the similar setting as Section \textbf{Sensitivity and Convergence Analysis (RQ4)} of the main paper.
We plot the training loss of GTL and OTL across timestamps in Figure~\ref{fig:training:individual}. We observe that the training loss continued to decrease on all timestamps for both OTL and GTL.
The observation supports that GTL and OTL could converge.
Furthermore, we provide the full version of the following Figures of the main paper.
\begin{itemize}
    \item Figure~\ref{fig:dataset:top-interactions:apd}, informing about the temporal dynamics of datasets, is the full version of Figure~\ref{fig:dataset:top-interactions}.
    \item Figure~\ref{fig:embedding:apd}, informing about the shift of item embeddings, is the full version of Figure~\ref{fig:embedding}.
\end{itemize}
As can be observed, the explanations regarding the analysis of GTL and OTL from the main paper still hold.

\end{document}